\documentclass[11pt]{article}
\usepackage{a4}
\usepackage{amsmath}
\usepackage{amsfonts}
\usepackage{latexsym}
\usepackage{epsfig}
\newcommand{\be}{\begin{equation}}
\newcommand{\ee}{\end{equation}}
\newcommand{\bea}{\begin{eqnarray}}
\newcommand{\eea}{\end{eqnarray}}

\renewcommand{\d}{\mathrm{d}}

\newcommand{\Ra}{\Rightarrow}

\DeclareMathSymbol{\mg}{\mathrel}{symbols}{"1D}

\newcommand{\ga}{\alpha}

\renewcommand{\gg}{\gamma}
\newcommand{\gd}{\delta}
\renewcommand{\ge}{\epsilon}
\newcommand{\gve}{\varepsilon}
\newcommand{\gf}{\phi}

\newcommand{\gx}{\xi}
\newcommand{\gm}{\mu}
\newcommand{\gn}{\nu}

\newcommand{\gl}{\lambda}
\newcommand{\gr}{\rho}

\newcommand{\gs}{\sigma}

\newcommand{\gz}{\zeta}
\newcommand{\gp}{\pi}
\newcommand{\gps}{\psi}
\newcommand{\get}{\eta}
\newcommand{\gch}{\chi}
\newcommand{\gG}{\Gamma}
\newcommand{\gD}{\Delta}
\newcommand{\gF}{\Phi}

\newcommand{\gL}{\Lambda}

\newcommand{\gO}{\Omega}

\newcommand{\cD}{{\cal D}}

\newcommand{\cM}{{\cal M}}

\newcommand{\tD}{{\tilde D}}

\newcommand{\tr}{\text{tr}}

\newcommand{\Id}{\text{\small 1}\hspace{-3.5pt}\text{1}}
\newcommand{\slashed}{\hspace{-1.1ex}/}
\newcommand{\Slashed}{\hspace{-1.3ex}/\hspace{.2ex}}

\newcommand{\ra}{\rightarrow}

\newcommand{\der}{\partial}

\newcommand{\inv}{^{-1}}
\newcommand{\dsp}{\displaystyle}

\newcommand{\labl}[1]{\label{#1}}
\newcommand{\half}{\frac 12 }

\newcommand{\beq}{\begin{equation}}
\newcommand{\eeq}{\end{equation}}
\newcommand{\barr}{\begin{array}}
\newcommand{\earr}{\end{array}}
\newcommand{\equ}[1]{\begin{gather} #1 \end{gather}}
\newcommand{\equa}[1]{\begin{align} #1 \end{align}}

\newcommand{\tabu}[2]{\begin{tabular}{#1} #2 \end{tabular}}
\newcommand{\arry}[2]{\begin{array}{#1} #2 \end{array}}

\newcommand{\mtrx}[1]{\begin{matrix} #1 \end{matrix}}
\newcommand{\pmtrx}[1]{\begin{pmatrix} #1 \end{pmatrix}}

\newcommand{\non}{\nonumber}
\newcounter{oldcounter}

\newcommand{\bgve}{{\bar\varepsilon}}
\newcommand{\bgf}{{\bar\phi}}

\newcommand{\bgz}{{\bar\zeta}}

\newcommand{\bgps}{{\bar\psi}}
\newcommand{\bget}{{\bar\eta}}
\newcommand{\bgch}{{\bar\chi}}
\newcommand{\tgf}{{\tilde\phi}}

\newcommand{\Natr}{\mathbb{N}}
\newcommand{\Intr}{\mathbb{Z}}

\newcommand{\sgn}{\text{sgn}}

\if@twoside\oddsidemargin25mm
\else
\begin{document}

\begin{flushright}
hep-th/0205012
\end{flushright}
\vskip 2 cm
\begin{center}
{\Large {\bf Instabilities of bulk fields and anomalies 
on orbifolds} 
}
\\[0pt]

\bigskip
\bigskip {\large
{\bf S.\ Groot Nibbelink ${}^{a,}$\footnote{
{{ {\ {\ {\ E-mail: nibblink@th.physik.uni-bonn.de}}}}}}}, 
{\bf H.P.\ Nilles ${}^{a,}$\footnote{
{{ {\ {\ {\ E-mail: nilles@th.physik.uni-bonn.de}}}}}}},
{\bf M.\ Olechowski ${}^{a,b,}$\footnote{
{{ {\ {\ {\ E-mail: olech@th.physik.uni-bonn.de}}}}}}}
\bigskip }\\[0pt]
\vspace{0.23cm}
{\it ${}^a$ Physikalisches Institut der Universit\"at Bonn,} \\
{\it Nussallee 12, 53115 Bonn, Germany.}\\
\vspace{0.23cm}
{\it ${}^b$ Institute of Theoretical Physics, Warsaw University,} \\
{\it Ho\.za 69, 00--681 Warsaw, Poland.}\\

\bigskip
\vspace{3.4cm} Abstract
\end{center}
Bulk matter modes of higher dimensional models generically become
unstable in the presence of additional matter multiplets at the
branes. This quantum instability is driven by localized
Fayet--Iliopoulos terms that attract the bulk zero modes towards the
boundary branes. We study this mechanism in the framework of a 5
dimensional $S^1/\Intr_2$ orbifold and give conditions for the various
possibilities of localization of (chiral) zero modes. This mechanism is
quite relevant for realistic model building, as the standard model
contains $U(1)$ hypercharge with potentially localized FI--terms. The
analysis is closely related to localized anomalies in higher
dimensional gauge theories. Five dimensional gauge invariance of the
effective action determines the anomaly constraints and fixes the
normalization of Chern--Simons terms. The localization of the bulk
modes does not effect the anomaly cancellation globally, but the
additional heavy Kaluza--Klein modes of the bulk fields may cancel
the Chern--Simons terms. 
We discuss also the potential appearance of the parity anomaly that 
might render the construction of some orbifold models inconsistent. 

\newpage

\section{Introduction}

There has recently been a large interest in field theory models of
extra dimensions with boundaries. This opens up the possibility 
that within a given scheme one might have fields living in various
dimensions, 
which may interact with each other in complicated though fascinating
ways. The ultimate picture is that such models are presumably part of a
low-energy description of string or M--theory, and for that reason it
is not 
surprising that some amount of supersymmetry is present in these extra
dimensions. As supersymmetry in higher dimensions corresponds to
extended supersymmetry in an effective 4 dimensional theory, some if
not all of those supersymmetries should be broken to give us a
phenomenologically interesting model that may incorporate the Standard
Model of particle physics. Boundaries of, or branes in, extra dimensions
lead naturally to a certain amount of supersymmetry breaking and often
also to a chiral spectrum in the effective 4 dimensional description. 

In the present paper we would like to study two important aspects of
theories in extra dimensions, namely their consistency and stability. 
We consider 5 dimensional supersymmetric field theories, with the
extra dimension compactified on the orbifold $S^1/\Intr_2$. The
orbifold condition naturally leads to boundaries that break
half of the supersymmetries. At those branes additional matter may
be situated that form well--defined representations of the remaining
supersymmetries. For simplicity we consider only the possibility that
chiral multiplets (containing chiral fermions and complex scalars) are
present on the branes. In the 5 dimensional bulk we assume that a set
of hyper multiplets (each consisting of two complex scalars and a
Dirac fermion), and vector multiplets (each with 5 dimensional
gauge field, a real scalar and a Dirac fermion) are situated. 

There can be potential instabilities in models with an arbitrary
distribution of matter fields over the bulk and the branes, which
indicate that some setups are extremely disfavored.   
In particular for a $U(1)$ super Yang-Mills--theory in the bulk, we
observed in  
\cite{GrootNibbelink:2002wv} that the divergent Fayet-Iliopoulos
(FI)--terms 
\cite{Ghilencea:2001bw}
localized at the branes  
\cite{Barbieri:2001cz,Scrucca:2002eb}
can lead to localization of the
zero modes of bulk hyper multiplets. (Localization has been considered
in the past, see for example \cite{jackiw1976}. Consequences of
Fayet-Iliopoulos 
terms on the shape of zero modes have also been studied at the purely
classical level in \cite{Arkani-Hamed:2002tb,Kaplan:2001ga}.) 
This dynamical mechanism tells
us that certain configurations are unstable, and may therefore not be
the appropriate starting point for a thorough phenomenological
investigation of such models. In this paper we take a closer look at
this phenomenon of localization: in particular we would like to
understand which forms of the localized modes are possible and under
which conditions they occur. 

An even more serious effect signaling instabilities, if not worse, of
a quantum theory are anomalies. 
In \cite{Arkani-Hamed:2001is} it was shown that anomalies in 5
dimensional theories can occur due to bulk fields, but they are
situated at the 4 dimensional boundaries only. (Using similar
methods anomalies on the orbifold $S^1/\Intr_2\times \Intr_2'$ have
been discussed in \cite{Scrucca:2002eb,Pilo:2002hu,Barbieri:2002ic}.)
One question that naturally comes up, is whether models with bulk and
brane matter can always be made consistent by an appropriate
Chern-Simons 
term. Even if this is possible, it seems that the consistency of the
low-energy spectrum depends on such high-energy interactions: this
goes somewhat against the intuition that for anomalies only the
low-energy spectrum (the zero modes) are relevant. 
In addition, does the localization of bulk fermions have consequences 
for the description of anomalies? In this work
we address these questions and show that they are related in an
interesting way.  

Another serious issue we raise in this paper concerns the very
definition of orbifold models with fermions: the orbifold
$S^1/\Intr_2$ is defined by dividing out a parity in the 5th dimension.
What happens if this parity is anomalous? We investigate whether 
a counter term can render the definition of a model to be consistent.

To achieve these goals we have organized the paper as follows:
In section \ref{sc:susy} we review properties of five dimensional
globally supersymmetric theories on $S^1/\Intr_2$. We consider the
following bulk multiplets: a $U(1)$ vector multiplet $V$ and
a set of charged hyper multiplets $H$ with charge operator $Q$. In
addition we allow for a set of chiral multiplets $C_0, C_\gp$ with
charge matrices $q_0, q_\gp$ on the branes at $y = 0$ and $y = \gp R$,
respectively. Following the arguments of ref.\ \cite{Mirabelli} the
induced brane supersymmetry transformations of bulk fields are
discussed. The consequences of this reasoning for possible FI--terms on
branes are confirmed  in section \ref{sc:tadpoles} by an explicit
calculation of the relevant tadpoles that may appear at one-loop. 
The resulting effective potential is investigated in section
\ref{sc:potential}. In particular we investigate the conditions under
which supersymmetric and $U(1)$ gauge symmetry preserving background
field configurations do exist. 

Next, we describe conditions for localization effects of zero modes to
occur in section \ref{s:localization}. 
A careful analysis is presented to classify the possible shapes of the
zero modes. This information is accumulated in table 
\ref{localPosibilities}. A discussion on the mass spectrum of the
non-zero modes concludes the section on localization. 
  
The final topic of this paper are anomalies. We first point out that 
the parity anomaly on the circle $S^1$ may complicate the
definition of the fermionic field theory on the orbifold
$S^1/\Intr_2$. After that, an alternative derivation of bulk field
anomalies is given using topological methods \cite{Horava:1996qa}. 
For consistency the local five dimensional gauge invariance of the 
orbifold model is essential, we work out the consequences for 
the matter content and interaction structure of such models. We show
that the localization of chiral zero modes changes the description of
the local anomalies, but do not affect the consistency of the
models.

\section{Bulk and (induced) brane supersymmetry}
\labl{sc:susy}

In the bulk we assume that one vector multiplet $V$ and a set of hyper
multiplets $H = \{H_b, b = 1, \ldots, n\}$ exist. We use  conventions
and normalizations from \cite{Zucker} for these multiplets, except the 
auxiliary fields for which we included a factor $-\half$ to comply
with the convention often used in the literature \cite{Mirabelli}. 
A $U(1)$ vector multiplet $V = (A_M, \gch, \gF, \vec D)$ in five
dimensions transforms under supersymmetry as
\equ{
\mtrx{
\gd A_M = i \bgve \gg_M \gch,
\qquad
\gd \gF = i \bgve \gch,
\qquad
\gd \vec D = \bgve \vec \gs \gg^M \der_M \gch, \\[2ex]
\gd \gch = \frac 14 F^{MN} \gg_{MN}\gve
- \half \gg^M \der_M \gF \gve -\half i \vec \gs \cdot \vec D \gve.
}
}
Here $A_M$ is a five dimensional gauge field, $F_{MN}$ is the
corresponding field strength.
The gaugino $\gch$ and the supersymmetry parameter $\gve$ are 
symplectic Majorana. As long as the $N=1$ supersymmetry in five
dimensions is manifest, it is convenient to use  symplectic
Majorana and related reality conditions. In appendix
\ref{a:conventions} we collected our conventions of notation and
give these reality conditions. To complete the description of the 
content of the vector multiplet, there is an iso--triplet of auxiliary 
scalars $\vec D$. The Lagrangian for the vector multiplet reads
\equ{
L_V = - \frac 14 F_{MN}F^{MN} + \half (\der_M\gF)^2
+ \half  \vec D^2 + i \bgch \gg^M D_M \gch.
\labl{eq:lagrV}
}
A hyper multiplet consists of a set of 4 real scalars $h$ (which are  
represented as a quaternion), a symplectic Majorana spinor $\gz$ 
(but with slightly different reality 
condition then the gauginos, see appendix \ref{a:conventions}), 
and a quaternion of 4 real auxiliary fields $F$.
The supersymmetry transformations of
the charge hyper multiplets $H = (h, \gz, F)$ take the form
\equ{
\mtrx{
\gd h = i \bgve \gz,
\qquad
\gd F = i \bgve \gg^M D_M \gz + 2 g_5 \bgve Q h \gch
+ g_5 \bgve \gF Q \gz,
\\[2ex]
\gd \gz = - \gg^M D_M h \gve - F \gve - i g_5 \gF Q h \gve,
\qquad
D_M = \der_M + i g_5 A_M Q,
}
}
with $g_5$ the five dimensional gauge coupling and $Q$ the
$U(1)$ charge operator. The Lagrangian for the hyper multiplets is
given by
\equ{
\mtrx{
L_H = \tr \left[
\frac 12 D_M h^\dag D^M h  + \half F^\dag F
- \half g_5^2 \gF^2 h^\dag Q^2 h
-\half  g_5 h^\dag Q h \vec \gs \cdot \vec D
\right]
\\[2ex]
+ \frac i2 \bgz \gg^M D_M \gz + \half g_5 \gF \bgz Q \gz
+ 2 g_5 \bgz Q h \gch.
}
\labl{eq:lagrH}
}
The trace here is over the $USp(2)$ indices.
As the bulk is the interval $S^1/\Intr_2$ times four dimensional
Minkowski space, orbifold boundary conditions have to be given.

Because of the orbifolding conditions for the fermions, it is not
possible to preserve the full five dimensional supersymmetry
structure (the 8 super charges). However, the orbifolding preserves 4
global super charges \cite{Bergshoeff}: 
\(
\gve = \gve(y) = \gve(-y) = \gs_3 \, i\gg^5\, \gve(y) = 
\gs_3 \, i\gg^5\, \gve. 
\)  
By requiring, in addition, that the gauge symmetry is unbroken, 
the orbifolding conditions are uniquely defined up to similarity
transformations by stating them for the fermions. Concretely, we
have for the boundary conditions
\equ{
\mtrx{
\gch(-y) = \gs_3 \, i\gg^5 \gch(y),
\\[2ex]
\vec \gs \cdot \vec D(-y) = \gs_3  \vec \gs \cdot \vec D(y) \gs_3,
}
\qquad
\mtrx{
h(-y) = - (\Id \otimes \gs_3) h(y) \gs_3,
\\[2ex]
F(-y) = (\Id \otimes \gs_3) F(y) \gs_3,
}
\qquad
\mtrx{
\gz(-y) =  (\Id \otimes \gs_3) i \gg^5 \gz(y),
\\[2ex]
\gF(-y) = - \gF(y),
}
\labl{eq:parity}
}
and charge operator is given by $Q = -q \otimes \gs_3$.

The orbifold conditions and the reality conditions can be ``solved''
simultaneously using the relations
\equ{
h = \pmtrx{ ~~\gf_-^* & \gf_+ \\ - \gf_+^* & \gf_-}, \quad
F = \pmtrx{ f_+ & - f_-^* \\ f_- & ~~f_+^*}, \quad
\gch =  \pmtrx{\gl~ \\ - \gl^{C_+}}, \quad 
\gz =\pmtrx{ \gps~ \\ - \gps^{C_+}}, \quad
\gve =\pmtrx{ \get~ \\ - \get^{C_+}}.
\labl{SolReal}
}
The Dirac fermion $\gl$ can be decomposed into chiral projection on
four dimensional Majorana spinors $\gl_\pm$ as
$\gl = \gl_{+L} - \gl_{-R}$ and $\gl^{C_+} = \gl_{+R} + \gl_{-L}$.
The subscript $\pm$ refers to the eigenvalues under the parity operator
$\gs_3 i\gg^5$.
The right-handed states are not independent from the left-handed ones:
 $(\gl_{-R})^{C_-} = \gl_{-L}$, etc..
The supersymmetry parameter $\gve$ and the hyperino $\gz$ can be
represented by four dimensional Majorana spinors in a similar
fashion. The parities of all fields in the model
are given by the following table
\equ{
V:~
\arry{l|c|c|c|c|c|c|c|}{
\text{state} &A_\gm & A_5 & \gF & \gl_{\pm L} & \gl_{\pm R} & D_3 & D_{1,2}
\\\hline
\text{parity} & + & - & - & \pm & \pm & + & -}
\qquad
H:~ \arry{l|c|c|c|c|}{
\text{state} &\gf_\pm & \gps_{\pm L} & \gps_{\pm R} & f_\pm
\\\hline
\text{parity} &\pm & \pm & \pm & \pm }
\non}
The supersymmetry parameter $\get_+$ corresponds to the 4 unbroken
supersymmetry charges ($\get_- =0$). 
As observed in \cite{Mirabelli}, the
transformations of the odd and even fields reside in different
multiplets with respect to this $N=\frac 12$ supersymmetry from the
five dimensional point of view. For the even fields the transformation
rules under this unbroken supersymmetry read 
\equ{
\mtrx{
\gd A_\gm = i \bget_+ \gg_\gm \gl_+,
\qquad
\gd \gl_+ = \frac 14 F^{\gm\gn} \gg_{\gm\gn} \get_+ -\half i \tD_3 \get_+,
\qquad
\gd \tD_3 =  \bget_+ \gg^\gm \der_\gm \gl_+,
\\[2ex]
\gd \gf_+ = i \bget_{+R} \gps_{+L},
\qquad
\gd \gps_{+L} = - \gg^\gm D_\gm \gf_+ \get_{+R}
- \tilde f_+ \get_{+L},
\\[2ex]
\gd \tilde f_+ = \bget_{+L}
( i \gg^\gm D_\gm \gps_{+L} + 2 g_5 q \gf_+ \gl_{+R} ),
}
}
but with ``modified'' auxiliary fields $\tD_3 = D_3 - \der_y \gF$
and $\tilde f_+ = f_+ - \der_y \gf_-^*$. For odd fields similar
transformations rules can be given.  However, as the even fields
(unlike the odd fields) do not vanish on the branes, they form $N=1$
four dimensional supersymmetric multiplets on the branes: 
a vector multiplet
$V| = (A_\gm, \gl_+, \tD_3)$ and chiral multiplets
$C_+| = (\gf_{+}, \gps_{+L}, \tilde f_{+})$.

In addition to these multiplets that are induced from the vector and
hyper multiplets in the bulk, we assume that there
are an arbitrary number of chiral multiplets
$C_0 = (\gf_0, \gps_{0L}, \tilde f_0)$ and
$C_\gp = (\gf_\gp, \gps_{\gp L}, \tilde f_\gp)$ on the branes at $y = 0$ and
$y = \gp R$, respectively. Their supersymmetry transformation rules
are identical to those of $C_+|$. Using the standard four dimensional
tensor calculus invariant Lagrangians can be obtained:
\equ{
\arry{rl}{
L_{branes} = {\dsp \sum_{I = 0, \gp}} \gd(y - I R) &
\left\{
D_\gm \gf_I^\dag D^\gm \gf_I + i \bgps_{IL} D\Slashed \gps_{IL}
+  f_I^\dag f_I +
\right.
\\[2ex] &
\left.
+ g_5 \gf_I^\dag q_I \gf_I (-D_3 + \der_y \gF)
-2 g_5 ( \bgps_{IL} \gl_{+R} q_I \gf_I + \text{h.c.})
\right\}.
}
\labl{eq:lagrbranes}
}

\section{Tadpoles}
\labl{sc:tadpoles}

In the previous section the (unbroken) supersymmetry transformations
for bulk vector (and hyper) multiplets have been discussed.
Not $D_3$ but $\tD_3 = D_3 - \der_y \gF$ is
the relevant auxiliary field for the remaining supersymmetry; this is
true in particular on the branes.  
The one-loop FI--terms, due to brane chiral multiplets $C_0$ or
$C_\gp$, are proportional to $\tD_3$; this is obvious as for them
$\tD_3$ is the 
relevant auxiliary component. (The non-renormalization of the 
FI--terms in four dimensions has been proven in \cite{Fischler}.)
If supersymmetry is preserved at the
one-loop level one would expect a similar result for the FI--tadpole
due to hyper multiplets in the bulk: again this tadpole should be
proportional to $\tD_3$. In ref.\ \cite{Ghilencea:2001bw} it was
shown that hyper multiplets may lead to a quadratically divergent
FI--term, and in refs.\ \cite{Barbieri:2001cz,Scrucca:2002eb} 
 it was argued that
the counter term for $D_3$ is located at the branes. For completeness
we redo a similar calculation here, but at the same time we show that
also $\der_y \gF$ counter terms are needed on the boundaries, as one
would expect from the supersymmetry analysis of the previous section.
In ref.\ \cite{GrootNibbelink:2002wv} the source of the divergence of
$\gF$ from bulk fields was identified to be the bulk fermions: the
hyperinos. 

Let us start with some more formal arguments to determine the generic
structure of possible tadpoles. Notice that the Lagrangian
\eqref{eq:lagrH} is quadratic in the hyper multiplet fields, 
hence the path integral over the hyper multiplets is
formally trivial to compute and gives rise to determinants.
Expanding to first order in $\gF$ and $\vec D$ this gives us
\equ{
\arry{lcl} {
i \gG_{eff} =
\ln \int \cD h \cD \gz e^{i S_H} & \approx &
\tr_\gz \ln \Bigl( \der\slashed - i \half \gF Q \Bigr) -
\tr_h \ln \Bigl( i \Box \otimes \Id  +\half i g_5 Q \otimes \vec \gs \cdot
\vec D \Bigr)
\\[2ex] &\approx &
- g_5 \tr_\gz \Bigl( \Box\inv i \der\slashed \gF Q \Bigr)
+ \half g_5 \tr_h \Bigl( \Box\inv Q \otimes \vec \gs \cdot \vec D \Bigr).
}
}
The zeroth order term cancels out because of supersymmetry.
The traces over the hyperons and hyperinos, of course, have to
be invariant under the boundary conditions \eqref{eq:parity}. This
shows that of the fermion and boson traces only
\(
\tr_\gz \Bigl( \Box\inv i \gg^5 \der_y \gF Q \Bigr)
\)
and
\(
 \tr_h \Bigl( \Box\inv Q D_3 \Bigr)
\)
survive, respectively. Notice that this in particular shows,
that the $D_1, D_2$ do not receive tadpole corrections. 
And that the field $\gF$ will only appear with a derivative acting on
it: $\der_y \gF$. 
Next, we compute those remaining traces using mode expansions.

As this calculation can be preformed using standard Feynman rules
if we employ Dirac spinors and complex scalars instead of symplectic
Majorana fermions and hyperons, we rewrite the relevant part of the
hyper multiplet Lagrangian as
\equ{
\arry{lcl}{
L_H & \supset &
D_M \gf_{+}^\dag D^M \gf_{+} + D_M \gf_{-}^\dag D^M \gf_{-}
- g_5 D_3 (\gf_{+}^\dag q \gf_{+} - \gf_{-}^\dag q \gf_{-})
\\[2ex] &&
+ i \bgps \gg^M D_M \gps
- g_5 \gF \bgps q \gps,
\\[2ex]
L_{branes} & \supset &
\gd(y) \left( D_\gm \gf_0^\dag D^\gm \gf_0
+ g_5 (- D_3 + \der_y \gF) \gf_0^\dag q_0 \gf_0
\right)
\\[2ex] & &
+  \gd(y - \gp R) \left(
D_\gm \gf_\gp^\dag D^\gm \gf_\gp
+ g_5 (- D_3 + \der_y \gF) \gf_\gp^\dag q_\gp \gf_\gp
\right).
}
}
(This shows that the original hyper multiplet was normalized such that
the standard normalization of complex bosons and Dirac fermions is
obtained.)

The mode expansions, into even and odd state w.r.t.\ the parity, of
the hyperons $\gf_\pm$, hyperinos $\gps_\pm$, 
the gauge scalar $\gF$ and the auxiliary field $D_3$ read
\equ{
\mtrx{
\gf_{+}(y)  = \get_n \cos \frac {ny}R \gf_{+}^n,
\\[2ex]
\gf_{-}(y)   = \get_n \sin \frac {ny}R \gf_{-}^n,
}
\qquad
\mtrx{
\gps_{+L}(y)   = \get_n \cos \frac {ny}R \gps_{+L}^n,
\\[2ex]
\gps_{-R}(y)   = \get_n \sin \frac {ny}R \gps_{-R}^n,
}
\qquad
\mtrx{
D_3(y)   = \get_n \cos \frac {ny}R D_n,
\\[2ex]
~\gF(y)   = \get_n \sin \frac {ny}R \gF_n,
}
}
where, to take care of the different normalization of the zero mode,
 $\get_0 = 1/\sqrt{\gp R}$ and $\get_{n>0} = \sqrt{2/\gp R}$.
In terms of these mode functions the propagators for the hyperons
$\gf^n_{\pm}$ and the hyperinos $\gps^n_{\pm}$ become
\equ{
< \gf_{+}^{n'\dag} \gf_{+}^{n''} >\inv\ =
\frac {\gd_{n' n''}}{p_4^2 - {n'}^2/R^2},
\qquad
< \bgps_{+}^{n'\dag} \gps_{-}^{n''} > \inv\ =
\frac {\gd_{n' n''}}{p_4^2 - {n'}^2/R^2}
\pmtrx{
p\slashed_4 R & - \frac {n''}R L \\
- \frac {n''}R R & p\slashed_4 L
},
\labl{Props}
}
respectively. And the vertices of $D_3$ with $\gf_\pm$ and
$\gF$ with $\gps_\pm$ decomposed in terms of these modes give 
\equ{
\mtrx{
< D_n \gf_{+}^{n'\dag} \gf_{+}^{n''} > &= &
\!\!\!\! - g_5 q {\dsp \frac {\gp R}4}
\get_n \get_{n'} \get_{n''}
\Bigl(
 \gd_{n',n+n''} + \gd_{n'',n+n'} + \gd_{n,n'+n''} + \gd_{n+n'+n'',0}
\Bigr),
\\[2ex]
< D_n \gf_{-}^{n'\dag} \gf_{-}^{n''} > &= &
g_5 q {\dsp \frac {\gp R}4}
\get_n \get_{n'} \get_{n''}
\Bigl(
 \gd_{n',n+n''} + \gd_{n'',n+n'} - \gd_{n,n'+n''} - \gd_{n+n'+n'',0}
\Bigr),
\\[3ex]
< \gF_n \bgps_{+L}^{n'\dag} \gps_{-R}^{n''} > &= &
~~g_5 q {\dsp \frac {\gp R}4}
\get_n \get_{n'} \get_{n''}
\Bigl(
- \gd_{n',n+n''} + \gd_{n'',n+n'} + \gd_{n,n'+n''} - \gd_{n+n'+n'',0}
\Bigr),
\\[2ex]
< \gF_n \bgps_{-R}^{n'\dag} \gps_{+L}^{n''} > &= &
~~g_5 q {\dsp \frac {\gp R}4}
\get_n \get_{n'} \get_{n''}
\Bigl(
+ \gd_{n',n+n''} - \gd_{n'',n+n'} + \gd_{n,n'+n''} - \gd_{n+n'+n'',0}
\Bigr),
}
}
respectively. The factor $\frac {\gp R}4$ arise by performing the
integral over the interval $[0, \gp R]$.  With these intermediate results 
the diagrams where the hyperons $\gf_\pm$ and the hyperinos 
$\gps$ run around in the loop 
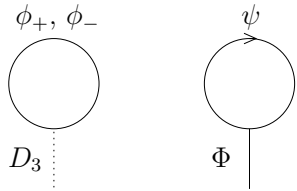
\begin{center}
\scalebox{.9}{\mbox{\input{FIbulk.pstex_t}}}
\end{center}
can be calculated. By putting the propagators for the bosons and
fermions together with the appropriate vertices, we find 
\equ{
\arry{rcl}{
\gx_{D_3} &=& 
\qquad 
{\dsp 
 g_5 q\, \frac {\gp R}{4} \,  \sum_{n,n',n''} 
\get_n \get_{n'} \get_{n''}\,  2 \gd_{n,n'+n''} \, 
{\dsp \int \frac {d^4 p_4}{(2 \gp)^4}}
\frac {\gd_{n',n''}}{p_4^2 - \frac {n^{\prime\, 2}}{R^2}} \,
D_{n} 
}
\\[2ex]
\gx_{\gF} ~ &=  & 
(-2)
{\dsp
 g_5 q\,  \frac {\gp R}{4}  \,\sum_{n,n',n''} 
\get_n \get_{n'} \get_{n''} \, 2 \gd_{n,n'+n''} \, 
{\dsp \int \frac {d^4 p_4}{(2 \gp)^4}}
\frac {\gd_{n',n''}}{p_4^2 - \frac {n^{\prime\, 2}}{R^2}} 
\, \Bigl( - \frac {n''}{R} \Bigr) \, 
\gF_{n}. 
}
}
}
Some comments are in order here: the factor $(-2)$ is due the trace
over chiral fermions. In particular, we 
have used here that because of the chiral projections in 
\eqref{Props} in the propagators only terms proportional to 
$n''/R$ survive. By using the consequences of the Kronecker deltas,
this can be represented as the derivate of $\gF$. Evaluating these
expressions further, gives the result
\equ{
\gx_{bulk} =
{\dsp \sum _{ n}}
g_5 \tr (q) {\dsp \int \frac {d^4 p_4}{(2 \gp)^4} \frac 1{p_4^2 - n^2/R^2}
}\get_{2n}
\Bigl(
- D_{2n} + (\der_y \gF)_{2n}
\Bigr). 
}
In this work we are only interested in the counter term structure of
the theory which is determined by the divergent parts of (one-)loop
results. 
Following \cite{Scrucca:2002eb} the divergent part of the remaining
four dimensional integral is regulated using the cut-off $\gL$ scheme
\equ{
\left. {\dsp \int \frac {d^4 p_4}{(2 \gp)^4} \frac 1{p_4^2 - n^2/R^2}}
\right|_{div}
= \frac {1}{16 \gp^2}
\Bigl(
\gL^2 - \frac {n^2}{R^2} \ln \gL^2
\Bigr).
}
This shows that there is a leading quadratic divergence, which
one would expect for Fayet-Iliopoulos tadpoles. But in addition
there is also a sub-leading logarithmic divergence, that needs to
be canceled by a counter terms as well. (Using dimensional
regularization one can derive similar conclusions provided one
introduces an additional infrared regulator to pick up the effects
due to zero-modes.) In terms of a coordinate space representation
of the divergent tadpoles, the $y$ dependent FI-parameter then reads
\equ{
\gx_{bulk}(y) = g_5 \frac{\tr(q)}{2}
\left(
\frac{\gL^2}{16 \gp^2} +
\frac{\ln \gL^2}{16 \gp^2} \frac 14 \der_y^2
\right)
\Bigl[ \gd(y) + \gd(y - \gp R) \Bigr].
}
We will see in later sections that the appearance of the double 
derivative of the delta--functions has important consequences. 
In addition to these bulk contributions, the diagrams 
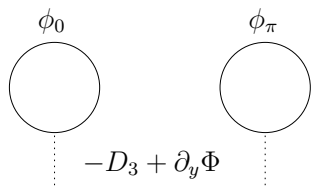
\begin{center}
\scalebox{.9}{\mbox{\input{FIbrane.pstex_t}}}
\end{center}
indicate that from the charged brane scalars we obtain 
the standard -- 4--dimensional -- FI--terms
\equ{
\gx_{branes}(y) = 
g_5  {\dsp \int \frac {d^4 p_4}{(2 \gp)^4} \frac 1{p_4^2}}
\sum_{I = 0, \gp}
\gd(y - I R)  \tr(q_I)
=
 g_5 {\dsp \frac {\gL^2}{16 \gp^2} }
\dsp 
\sum_{I = 0, \gp}
\gd(y - I R)  \tr(q_I). 
}
This together with the bulk FI--terms gives the full FI--parameter
\equ{
\gx(y) = \gx_{bulk}(y) + \gx_{branes}(y).
\labl{TotalFIy}
}

\section{Effective four dimensional potential}
\labl{sc:potential}

An analysis of the effective potential of this model is discussed
next. With the notion ``effective potential'' is meant the part of
the Lagrangian that can acquire Vacuum Expectation Values (VEVs)
that do no not break four dimensional Poincar\'e invariance.
This implies that we allow for background solutions that are functions of $y$.
As our analysis of this effective potential is essentially classical,
we just assume that there are FI--terms on the two boundaries
\equ{
L_{FI} =
\Bigl( - D_3 + \der_y \gF \Bigr) 
\sum_{I=0,\gp} 
\Bigl( \gx_I + \gx_I^{\prime\prime} \der_y^2 \Bigr)
\gd(y - IR),
\labl{eq:lagrFI}
}
with values $\gx_0, \gx_0^{\prime\prime}$ and $\gx_\gp,
\gx_\gp^{\prime\prime}$, respectively. The double derivative of the
delta--functions was not considered in the classical FI--term 
analysis of refs.\ \cite{Arkani-Hamed:2002tb,Kaplan:2001ga}. 
If we assume that the FI--terms are dominated by the one-loop induced
values we see that
\equ{
\gx_I = g_5 \frac{\gL^2}{16 \gp^2}
\Bigl(
\half \tr(q) + \tr (q_I)
\Bigr),
\qquad
\gx_I^{\prime\prime} = \frac 14 g_5 \frac{\ln \gL^2}{16 \gp^2}
\half \tr(q).
}
Notice, in particular, that 
$\gx_0^{\prime\prime} = \gx_\gp^{\prime\prime}$  
when they are generated at the one-loop level. The FI--parameters
$\gx_0$ and $\gx_\gp$ receive contributions from both fields in the
bulk and on the respective brane, while 
$\gx_0^{\prime\prime}= \gx_\pi^{\prime\prime}$ only
receives bulk contributions. Therefore the one-loop FI--term is
completely absent, if the sum of the charges in the bulk and on both
branes vanish separately. 

By simply collecting
the various relevant terms from the Lagrangians $L_V$, $L_H$,
$L_{brane}$ and $L_{FI}$ an expression for the effective potential is
obtained that is not manifestly semi-positive. 
However, by adding and subtracting 
the term $\half \der_y \gF \tr [h^\dag Q h \gs_3]$ and by doing an
integration by parts on the subtracted one and using that the
boundary terms are absent since $\gF$ is odd, the effective potential
takes the form
\equ{
\arry{rl}{
V_{4D} =  {\dsp \int_0^{\gp R} \d y}& \Bigl\{
\half \Bigl(~ \der_y \gF - \gx + \half g_5 \tr [h^\dag Q h \gs_3]
+ \gd(y)\gf_0^\dag q_0 \gf_0
+ \gd(y-\gp R) \gf_\gp^\dag q_\gp \gf_\gp\Bigr)^2
\\[2ex] &
- \half \Bigl( - D_3 - \gx + \half g_5 \tr [h^\dag Q h \gs_3]
+ \gd(y)\gf_0^\dag q_0 \gf_0
+ \gd(y-\gp R) \gf_\gp^\dag q_\gp \gf_\gp
\Bigr)^2
\\[2ex] & 
+ \frac 18 g_5^2 {\dsp \sum_{i=1,2}}
\Bigl( \tr [ h^\dag Q h \gs_i  ] \Bigr)^2
- \half {\dsp \sum_{i=1,2}}
\Bigl( -D_i + \half g_5 \tr[h^\dag Q h \gs_i] \Bigr)^2 +
\Bigr.
\\[2ex] &
\Bigl.
+ \half \tr \Bigl[
\bigl(
\der_y h^\dag - i g_5 A_5 h^\dag Q - g_5 \gF \gs_3 h^\dag Q
\bigr)
\bigl(
\der_y h + i g_5 A_5 Q h - g_5 \gF Q h \gs_3
\bigr)
\Bigr]
\Bigr\}.
}
}
At first sight, it might seem that this
expression is ill--defined because of the appearance of delta-functions
squared. However, working
out the squares shows that all (dangerous) delta-functions squared, in
fact, drop out. Furthermore this expression is manifestly semi-positive:
the negative squares that appear in the potential are zero using the
equations of motions for $D_3$ and $D_i$.

Before we study supersymmetric background solutions in detail, let us
say a few words about an important general property of background 
solutions on the boundaries. The vacuum field equations for
$\gf_0^\dag$ and $\gf_\gp^\dag$ are trivial to determine:
\equ{
q_0 \left< \gf_0 \right>  \left<-D_3 + \der_y \gF\right>(0) = 0, 
\qquad 
q_\gp \left< \gf_\gp \right>  \left<-D_3 + \der_y \gF\right>(\gp R) = 0.
}
If none of the states in $\left<\gf_0\right>$ or $\left< \gf_\gp\right>$
are chargeless, these equations imply that either supersymmetry 
is unbroken on the branes or the $U(1)$ symmetry is unbroken.

\subsection{Supersymmetric background solutions}

Supersymmetric vacuum solutions form a special class of all possible 
solutions. As usual
their structure is easier to analyse than that of arbitrary (not
necessarily supersymmetric) solutions. The equations that have to
be satisfied in such a case are \equ{ \mtrx{ 
D_3 = \der_y \gF =
\half g_5 \tr [ h^\dag Q h \gs_3] + \gd(y) \gf_0^\dag q_0 \gf_0  
+ \gd(y-\gp R) \gf_\gp^\dag q_\gp \gf_\gp + \gx(y), 
\\[2ex]
\tr [ h^\dag Q h \gs_i] = 0, \qquad \der_y h + i g_5 A_5 Q h -
g_5 \gF Q h \gs_3 = 0. 
} }
For our subsequent discussion it is more convenient to leave the
manifest 5 dimensional $N=1$ supersymmetric notation, since it
will turn out that the bosonic components, $\gf_+$
and $\gf_-$, have entirely different behavior. By substituting
the solution \eqref{SolReal} of the reality condition for the
hyperon $h$, these relations become
\equ{
D_3 = \der_y
\gF = g_5( \gf_+^\dag q \gf_+ - \gf_-^\dag q \gf_-)
  + \gd(y) g_5 \gf_0^\dag q_0 \gf_0 + \gd(y-\gp R) g_5\gf_\gp^\dag
q_\gp \gf_\gp + \gx(y),
\labl{BPS1}
}\equ{
\gf_+^T q \gf_- = 0, \quad 
\der_y \gf_\pm \mp g_5 q \gF \gf_\pm = 0.
\labl{BPS2}
}
In addition, $A_5$ has been put to zero. This can be achieved by a
gauge transformation, since it vanishes on both boundaries already. 

For a supersymmetric background solution these equations 
have to be satisfied simultaneously, and in addition the fields 
have to satisfy the appropriate boundary conditions. These equations
show that four dimensional supersymmetry only requires that 
$D_3 - \der_y \gF$ vanishes, but not $D_3$ as one might expect. (This
would be required if the five dimensional supersymmetry is to be
preserved, but this is of course not possible while enforcing the
orbifold conditions, in general). Non-vanishing odd fields are often
not allowed because they are in conflict with the boundary conditions.

For the parity-odd bulk scalars $\gf_-$ the boundary condition
implies that their VEVs  are zero: $\left< \gf_- \right> = 0$. 
This can be seen by the following argument: The solution in the bulk 
for $\gf_-$ is easily found. But since $\gf_-$ is odd, the 
integral over its derivative vanishes on $[0, \gp R]$. From this it 
follows that 
\equ{
0 = \int_0^{\gp R} \d y \, \der_y \left< \gf_- \right> = 
\exp \left\{ - g_5 \int_0^{\gp R} \d y\left< \gF\right>  q \right\} 
\left< \gf_{0-} \right>, 
\labl{vanGFmin}
}
with $\left< \gf_{0-} \right>$ the integration constant. 
Hence, unless the integral over $\left<\gF\right>$ diverges, the integration 
constant has to be zero. In particular, this shows that the equations
involving $\gf_-$ in the supersymmetry conditions \eqref{BPS2} 
are fulfilled trivially: $\left< \gf_{0-} \right> = 0$.  

The scalar of the gauge multiplet $\gF$ is odd as well, it vanishes on
the boundaries, and this leads to the integrability condition
\equ{
 0 =  \int_0^{\gp R}  \!\!\!\! \d y\, \der_y \gF 
= \int_0^{\gp R} \!\!\!\! \d y\, \left< \gf_{0+} \right>^\dag 
g_5 q \exp \left\{ 2 g_5 \int_0^{y}\!\!\! \d y\left< \gF\right>  q  \right\} 
\left< \gf_{0+} \right> 
+ \half \sum_I 
\left( \left< \gf_I\right>^\dag g_5 q_I \left< \gf_I\right>   + \gx_I
\right).
\labl{SusyGauge}
}
Here we have used that $\left< \gf_- \right> = 0$. There is no
contribution proportional to $\gx_I^{\prime\prime}$ since the 
first derivative of the delta function vanishes on the boundaries.  
Only the zero mode 
$D_{3\, 0}= \frac 1{\gp R}\int_0^{\gp R} \d y\, D_3 $
of the auxiliary field $D_3$ is relevant for the question whether 
supersymmetry is broken due to gauge interactions (D-terms).

Let us analyze under what conditions this necessary (though in general not
sufficient) requirement \eqref{SusyGauge} for supersymmetric 
background is satisfied for the two distinct cases: broken or unbroken
gauge symmetry. 
Suppose first that a background that does not spontaneously break the 
$U(1)$ gauge symmetry is possible. In that case all charged scalars 
(in the bulk and on the branes) have to vanish: 
$\left< \gf_{+} \right>= 0$ and $\left< \gf_{I} \right> = 0$. 
This is only possible if the sum of the brane FI--parameters vanishes:
\equ{
\gx_{tot} = \gx_0 + \gx_\gp = 0 
\qquad \Rightarrow \qquad
\tr (q) + \tr (q_0) + \tr (q_\gp) = 0.
\labl{vanishingFI}
}
This gauge invariant background is unique if all charges have the same
sign. Otherwise, there can be many
flat directions which spontaneously break the gauge symmetry. 
The relation of the sum of charges follow upon assuming that the 
FI--terms are induced at the one-loop level. When the FI--terms are 
generated by such quantum effects, it is not possible that all
charges have the same sign since then the sum of the charges
in \eqref{vanishingFI} cannot be zero. Therefore, the one-loop
FI--terms have flat directions that break gauge symmetry
spontaneously. 

If the sum of the FI--contributions 
$\gx_{tot} = \gx_0 + \gx_\gp$ is not equal zero, then supersymmetry
can be maintained (i.e.\ eq.\ \eqref{SusyGauge} is satisfied), 
provided that there is at least one (brane or bulk) field with a
charge which has the sign opposite to $\gx_{tot}$. Hence in this 
case the gauge symmetry is always spontaneously broken. 
The gauge symmetry can then be broken by bulk or brane fields. 
Hence we see a similar situation then in 4 dimensions, if the (total 
integrated) FI--contribution is non-zero then either supersymmetry or 
gauge symmetry is spontaneously broken.

\section{Localization of charged bulk fields}
\labl{s:localization}

As we have observed in \cite{GrootNibbelink:2002wv} the divergent 
FI--terms on the branes signal that the theory has a severe 
instability, that can lead to localization. 
The next topic we want to investigate in the paper is
under what conditions charged bulk field can become localized and what
the form of this localization is. In the first subsection, we answer
when localization is certainly not possible, i.e.\ when the zero mode
is constant. 
This is important because it gives us a criterion to decide
what kind of bulk configuration are stable under quantum corrections.
It occurs that the constant zero mode can be stable only in a special
setup. 
Next, we investigate the possible shapes of the zero modes in the
second subsection. The analysis that is performed here, only applies
when we have a background that does not spontaneously break the 
gauge invariance. (The analysis can be extended to that case, but then 
not all even fields have zero modes due to the Higgs mechanism.) 
The main results of this analysis are summarized in 
tables \ref{localPosibilitiesCutFinite} and \ref{localPosibilities}.
We end this section with a discussion of the massive part of the 
spectrum.

\subsection{Conditions for constant zero mode}

Because we only consider vacua that do not spontaneously break the
gauge invariance, the relevant part Lagrangian which describes the
bulk scalars is quadratic in them, it
follows that the equation that determines the shape of the zero-mode
is the same as the equation that determines the background solution. 
(Of course the zero mode has to be properly normalized.)  As was argued, 
using eq.\ \eqref{vanGFmin}, the odd bulk fields can never have a 
non-trivial background solution. Hence there are no massless odd
zero-modes. 

The shape of the zero mode of a single even field\footnote{Notice
that we slightly abuse notation here since $\gf_+$ denotes a vector of
scalars in general, while now we refer to one component with charge
$q_b$. By the subscript $b$ we indicate that we really mean one
component, and not the whole vector.}
$\gf_+$ with charge $q_b$ can formally be stated as 
\equ{
\der_y \gf_{0+} -  g_5 q_b \left< \gF\right> \gf_{0+} = 0 
\qquad \Ra \qquad 
\gf_{0+}(y) = \exp\left\{  g_5 q_b \int_0^y \d y \left< \gF \right>  \right\} 
\bgf_{0+},
\labl{zeroMode}
}
and is therefore entirely determined by $\left< \gF\right>$. 
The constants $\bgf_{0+}$ have to be chosen such that the zero 
mode is normalized as 
\equ{
\int_0^{\gp R} \d y\,  |\gf_{0+}(y)|^2 = 1.
\labl{zeroNormal}
}
From \eqref{zeroMode} it is clear that for $\left<\gF\right> (y) = 0$,
the zero mode is constant over the 5th dimension.  

Before investigating the properties of such localized zero modes further,
we first turn to the question what the conditions are for the zero
modes to be localized at all. In fact, it is easier first to answer
when the zero mode does not get localized. Since $\left< \gF \right>$
must vanish for all $y$ in that case, also its derivative vanishes. But
then, since eq.\ \eqref{BPS1} has to hold for all $y$, it follows that
the conditions 
\equ{
\gx_I = 0, 
\qquad \qquad 
\gx_I^{\prime\prime} = 0,
\labl{nonLocalization}
}
have to be satisfied simultaneously.  (The reason is that the delta
functions, their derivatives and the zero modes are independent
'functions'  (distributions) on the interval.) 
With the assumption that the FI--terms are only one-loop induced,
these conditions given above become 
\equ{
\tr (q_I)  = 0, 
\qquad \qquad 
\tr (q) = 0. 
\labl{nonLocalFI}
}
The importance of these relations is that they indicate which
configuration of supersymmetric bulk and brane fields are stable under
quantum corrections. As we will see in the next subsections, if these
conditions are not satisfied the description of the theory at the quantum
level may be very different from the initial set-up, due to
localization effects.

\subsection{Shape of the zero mode}
\labl{s:shapeZero}

In this subsection we determine the profile of the scalar zero mode. 
As the model we consider is supersymmetric, it is to be expected that
the chiral fermionic zero mode has the same shape over the extra
dimension. This can also be seen directly from its equation of 
motion 
\equ{
\left( i \gg^5 \der_y - g_5 q_b \left< \gF \right> \right) \gps_{0+L} =
0:
\labl{ChiralZeroMode}
}
upon using that the eigenvalue of the chirality operator $i\gg^5$ on
the chiral state $\gps_{0+L}$ is $+1$, this equation becomes identical
to the scalar zero mode equation \eqref{zeroMode}. Therefore, we
now only focus on the scalar zero mode, knowing that all our
conclusions apply to the chiral zero mode as well. We will make 
use of this fact when we discuss anomalies in section
\ref{s:AnomLocal}. At the end of this subsection we
consider the limit $| \gx_0 | \ra \infty$ of our results,
corresponding to taking the cut-off to infinity.

The extent of the localization of the zero mode \eqref{zeroMode} is
determined by the solution of the equation for $\left< \gF\right>$ 
for a gauge invariant background 
($ \left< \gf_{0+} \right> = 0$ and $  \left< \gf_I \right> =0$) 
given by 
\equ{
\der_y \left< \gF \right> = 
\Bigl(\gx_0 + \gx_0^{\prime\prime} \der_y^2 \Bigr) \gd(y) 
+  \Bigl(-\gx_0 + \gx_0^{\prime\prime} \der_y^2 \Bigr) \gd(y - \gp R). 
}
Here we have used that 
$\gx_0^{\prime\prime} = \gx_\pi^{\prime\prime}$, 
and that only for $\gx_0 = - \gx_\pi$ the integrability
condition \eqref{vanishingFI} can be satisfied for a gauge and
supersymmetric background. 
The solution of this equation is obtained straightforwardly 
\equ{
\left< \gF\right> (y) = \half\, \gx_0\,  \sgn(y) + \gx_0^{\prime\prime} 
\Bigl( 
\gd'(y) + \gd'(y - \gp R)
\Bigr), 
\qquad 
\sgn(y) = \begin{cases}
~~~1 & \,~~0 < y < \gp R, \\
\,~~ 0 & \qquad ~~  y=0, \gp R, \\ 
-1 & \gp R < y < 2\gp R. 
\end{cases}
\labl{gFSusy}
}
And its integral
\equ{
\int_0^{y} \d y\, \left< \gF\right> (y) = \half \gx_0 
\Bigl( \gp R - | y - \gp R| \Bigr) + \gx_0^{\prime\prime} 
\Bigl( \gd(y) + \gd(y - \gp R) \Bigr)
\labl{intgF}
}
can be used to determine the shape of the zero modes in the 
presence of this special background. For this purpose we
have to show 
explicitly what we obtain, when we formally insert the integral of 
$\gF$, eq.\ \eqref{intgF} into the form of the zero mode 
\eqref{zeroMode}. The main problems here are: how to
interpret the delta-functions in the exponential, and how to take the
normalization condition into account. 

To perform explicit calculations we need to regularize the
delta-function in a suitable 
way.\footnote{In a previous publication \cite{GrootNibbelink:2002wv} 
we used a Gaussian regularization of the delta--function, to clarify
the localization effects. To be able to perform all integrations
explicitly we have chosen a different regularized form of the
delta--function in this section.} 
To this end the simple step-function is sufficient 
\equ{
\gd(y)  \ra \gd_\gr(y) = 
\begin{cases}
\frac 1{2\gr} & | y | < \gr, 
\\[2ex] 
0 & | y | > \gr. 
\end{cases}
}
It reproduces the delta-function in the limit $\gr \ra 0$. 
As this is a very simple prescription of the delta-function, the
normalization constant $\bgf_{\gr+}$  of zero mode 
\(
\gf_{\gr +} (y) = \exp 
\left\{ g_5 q_b \int_0^y \d y\,\left< \gF \right>\right\}
\bgf_{\gr +}
\)
can be computed explicitly
\equ{
\bgf_{\gr+}^{-2} = 
e^{g_5 q_b \gx_0^{\prime\prime}/\gr} 
\left[
\frac {e^{g_5 q_b \gx_0 \, \gr} -1}{g_5 q_b \gx_0} 
+ e^{g_5 q_b \gx_0 \, \gp R} \,
\frac {1-e^{-g_5 q_b \gx_0 \, \gr} }{g_5 q_b \gx_0} 
\right]
+ 
\frac {e^{g_5 q_b \gx_0 (\gp R -\gr)} - e^{g_5 q_b \gx_0 \, \gr}}
{g_5  q_b \gx_0}. 
\labl{zeroNormExp}
}
Having determined the normalization factor with this regularization of
the delta-function, next we investigate to what function the
regularized zero mode $\gf_{\gr+}^2$ tends in the limit that
$\gr \ra 0$.  Since the behavior of this limit strongly depends on the
(relative) signs of various parameters, we now distinguish the following 
three cases:  $\gx_0^{\prime\prime} q_b > 0$,
$\gx_0^{\prime\prime} q_b < 0$, and $\gx_0^{\prime\prime} = 0$. 
In the first case, where the charge $q_b$ and the FI--parameter
$\gx_0^{\prime\prime}$ have the same sign, the limit $\gr \ra 0$ we
obtain for the square of the zero mode is 
\equ{
\gf_{0+}^2(y) = 
\frac {2 e^{g_5 q_b \gx_0 \, y}}{1 + e^{g_5 q_b \gx_0 \, \gp R}}
\left[
\gd(y) + \gd(y-\gp R)
\right], 
\qquad \qquad 
\gx_0^{\prime\prime} q_b > 0.
\labl{wave>}
}
Hence the zero mode has the delta function support on the two 
fixed points, but the height at these two fixed points is not the
same. In the second case, we find that the zero mode does not live on
the boundaries of the interval 
\equ{
\gf_{0+}^2(y) = 
\frac {g_5 q_b \gx_0\,  e^{g_5 q_b \gx_0 \, y}}{e^{g_5 q_b \gx_0 \, \gp R}-1}
\begin{cases}
1 & 0 < y < \gp R, \\[2ex]
0 & y = 0, \gp R,
\end{cases}
\qquad \qquad 
\gx_0^{\prime\prime} q_b < 0.
\labl{wave<}
}
This shows that the shape in this case is somewhat peculiar: the zero
mode vanishes at both branes identically, while having an exponential
behavior on the open interval $]0, \gp R[$. In table
\ref{localPosibilitiesCutFinite} we have schematically drawn both
possible forms of localization. 
Notice that in both results the FI--parameter $\gx_0^{\prime\prime}$ does
not appear anymore, since apart from its sign it can be completely
absorbed in a rescaling of $\gr$ which tends to zero anyway. 

\begin{table}
\begin{center} 
\renewcommand{\arraystretch}{1.5}
\tabu{ c | c }{
 $\gx_0'' q_b > 0$ & $\gx_0'' q_b < 0$ 
\\[2ex] \hline &\\[-1ex]
~~~~~~
\raisebox{-2mm}{\scalebox{.9}{\mbox{\input{loc0pi2.pstex_t}}}}
~~~~~~
& 
~~~~~~
\raisebox{-2mm}{\scalebox{.9}{\mbox{\input{exp0pi.pstex_t}}}}
~~~~~~
%\\[-1ex] &\\ \hline 
}
\end{center}
\caption{The two basic shapes (eqs.\ \eqref{wave>} and \eqref{wave<})
of the zero mode with charge $q_b$ are displayed for a finite value of
the cut-off $\gL$. (In table \ref{localPosibilities} the possible
shapes are given in the infinite cut-off limit.)
Delta function localizations, denoted by the arrows, 
happens if the signs of the FI--parameter $\gx_0''$ and charge $q_b$ are
the same. When the signs are opposite, the wave function falls off 
exponentially, and in addition vanishes at both branes. 
To which of the two branes the zero mode gets more localized is
decided by the sign of $(\gx_0 - \gx_\gp) q_b$.
}
\label{localPosibilitiesCutFinite}
\end{table}
 
This rescaling is meaningless if $\gx_0^{\prime\prime}$ vanishes
identically, and the zero mode then takes the form  
\equ{
\gf_{0+}^2(y) = 
\frac {g_5 q_b \gx_0\,  e^{g_5 q_b \gx_0 \, y}}{e^{g_5 q_b \gx_0 \, \gp R}-1},
\qquad \qquad \qquad \qquad \qquad \qquad 
\gx_0^{\prime\prime} = 0. 
}
A similar case has been studied before in refs.\
\cite{Kaplan:2001ga,Arkani-Hamed:2002tb}.

Another intriguing possibility is when $\gx_0 = 0$. 
Eq.\ \eqref{wave>} then tells us that for 
$\gx_0^{\prime\prime} q_b >0$ the zero mode becomes a state that has
two delta-function supports on both branes
\equ{
\gf_{0+}^2(y) = 
\gd(y) + \gd(y-\gp R),
\qquad \qquad 
\gx_0 = 0, \qquad \qquad \gx_0'' q_b> 0.
\labl{Vangx>}
}
This is a dynamically generated state that has the property that it is
localized on two (widely separated) branes. 
For the opposite case 
$\gx_0^{\prime\prime} q_b<0$ eq.\ \eqref{wave<} leads to a zero mode 
that is constant over the interior of the bulk but vanishes at both
boundaries:
\equ{
\gf_{0+}^2(y) = 
\begin{cases}
1 & 0 < y < \gp R, \\[2ex]
0 & y = 0, \gp R,
\end{cases}
\qquad \qquad \gx_0 = 0, 
\qquad \qquad 
\gx_0^{\prime\prime} q_b < 0.
\labl{Vangx<}
}
This state is not localized: it is constant over the bulk, but as it
vanishes at the fixed points it cannot interact with brane fields.

When $\gx_0 \neq 0$ we can study what happens in the limit 
$|\gx_0| \ra \infty$. For the case $\gx_0'' q_b \geq 0$ this limit leads to
zero modes that are localized on one of the two branes
\equ{
\gf_{0+}^2(y) = 
\begin{cases}
2 \gd(y - \gp R) & \qquad\qquad  \gx_0 q_b > 0,  \\[2ex]
2 \gd(y) & \qquad\qquad  \gx_0 q_b < 0,
\end{cases}
\qquad\qquad 
\gx_0^{\prime\prime} q_b \geq 0.
\labl{limit>}
}
The sign of $\gx_0 q_b$ decides on which of the two branes the state get
localized. When $\gx_0'' q_b < 0$, we get an even more intriguing
result: 
 the zero mode is localized infinitely close to one of the fixed
points; but on this fixed point itself the zero mode vanishes
\equ{
\gf_{0+}^2(y) = 
\begin{cases}
2 \gd(y - \gp R + 0) & \qquad \gx_0 q_b > 0,  \\[2ex]
2 \gd(y - 0) & \qquad \gx_0 q_b < 0,
\end{cases}
\qquad\qquad 
\gx_0^{\prime\prime} q_b < 0.
\labl{limit<}
}
With the notation $\pm 0$ we indicate that this zero mode is localized
infinitely close, but not on a fixed point. 
From eqs.\ \eqref{limit>} and \eqref{limit<}, one sees that the
zero mode always gets localized at one of the two branes for very 
large $| \gx_0 |$ irrespectively of the other FI--parameter
$\gx_0^{\prime\prime}$. However, for $\gx_0^{\prime\prime} q_b<0$ the
zero mode does not actually end up on one of the two branes, but gets
localized infinitely close to it. 

\begin{table}
\begin{center}
\equ{
\gx_0'' \sim  \ln (\gL^2)\  \tr(q),
\qquad \quad 
\mtrx{
\gx_0  \sim \gL^2\  \bigl(\tr(q_0) + \half \tr(q) \bigr),
 \\[2ex]
\gx_\gp \sim \gL^2\ \bigl( \tr(q_\gp) + \half \tr(q) \bigr),
}
\qquad \quad 
\mtrx{\text{SUSY}  +  U(1)} 
\quad \Ra\quad  
\gx_0  = - \gx_\gp
\non
}
\renewcommand{\arraystretch}{1.5}
\tabu{l | c | c }{
$q_b \neq 0$ & $\gx_0'' q_b \geq 0$ & $\gx_0'' q_b < 0$ 
\\[2ex] \hline &&\\[-1ex]
\raisebox{10mm}{$\mtrx{\gx_0 q_b < 0 \\ (\gx_\gp q_b >0)}$}~~ & 
~~~~~~
\raisebox{-2mm}{\scalebox{.9}{\mbox{\input{loc0.pstex_t}}}}
~~~~~~
& 
~~~~~~
\raisebox{-2mm}{\scalebox{.9}{\mbox{\input{near0.pstex_t}}}}
~~~~~~
\\[-1ex] &&\\ \hline &&\\[-1ex]
\raisebox{10mm}{$\mtrx{\gx_\gp q_b < 0 \\ (\gx_0 q_b > 0)}$} & 
\raisebox{-2mm}{\scalebox{.9}{\mbox{\input{locpi.pstex_t}}}}
& 
\raisebox{-2mm}{\scalebox{.9}{\mbox{\input{nearpi.pstex_t}}}}
 \\[-1ex] &&\\ \hline &&\\[-1ex]
\raisebox{10mm}{$\mtrx{\gx_0= \gx_\gp = 0 \\ \gx_0'' \neq 0}$} & 
\raisebox{-2mm}{\scalebox{.9}{\mbox{\input{loc0pi.pstex_t}}}}
& 
\raisebox{-2mm}{\scalebox{.9}{\mbox{\input{cons0pi.pstex_t}}}}
\\[-1ex] &&\\ \hline 
}
\end{center}
\caption{This table schematically displays the different shapes of 
a (bulk) zero mode with charge $q_b \neq 0$, summarizing eqs.\
\eqref{limit>},  \eqref{limit<}, \eqref{Vangx>}, and \eqref{Vangx<} 
with the cut-off $\gL \ra \infty$.
This mode localizes at the brane with the sign of its FI--parameter  
($\gx_0, \gx_\gp$)  opposite to the charge of the zero mode. 
To allow for a supersymmetric vacuum solution that preserves the
$U(1)$ symmetry, these two FI--parameters add up to zero. Assuming 
that the one-loop FI--terms dominate, these FI--parameters are
proportional to the sum of charges at the branes, and the third
FI--parameter $\gx_0''$ is proportional to the sum of charges in the
bulk. 
If the product of $q_b$ and $\gx_0''$ is negative the zero mode localizes
infinitely close to but not at the brane, otherwise it localizes
exactly on the brane. Vanishing FI--terms $\gx_0, \gx_\gp$ is special: 
if $\gx_0'' q_b > 0$ the zero mode 
splits up into two delta functions on both branes,
while for $\gx_0'' q_b < 0$ the zero mode is constant over the
interior of the bulk but vanishes at its boundaries. 
When all FI--parameters vanish, of course, there is no localization effect. 
}
\label{localPosibilities}
\end{table}
 
In table \ref{localPosibilities} we have schematically drawn the
various possibilities for the localization of the zero mode, depending
on the charge of the bulk field and the signs of the brane
FI--parameters. In these pictures we have taken the limit $|\gx_0| \ra
\infty$ when it is not zero. 

The behavior of the bulk zero modes discussed in this 
subsection, can be viewed as a supersymmetric field theory analog of 
electrostatics. The system
tries to distribute its free charges such that everywhere the net
charge is zero. In this sense the configurations under investigation
here are very similar to having two charged plates with some fluid in
between which consists of charged particles as well. Also here the
charge particles of the fluid distribute themselves, so as the net
charge vanish everywhere if possible. This is because of the electric
field generated by the differences of the charges of the two plates.

\subsection{Mass spectrum}
\labl{s:MassSpec}

In the previous section we have seen that the scalar and chiral zero
modes get localized on (one of) the branes. 
But these are only just the lightest
modes of an infinite tower of states. Here we describe what happens to
the masses of the other modes in the Kaluza-Klein tower for a finite
FI--parameter $\gx_0$, so that we clearly see what happens in the
limit when this parameter tends to infinity. 

As for the zero modes, we restrict our discussion of the mass spectrum
to the scalars only. The masses for the fermions are identical, because 
their field equations can be ``squared'' to reproduce the boundary
value problem of the scalars. 

The spectrum of the bosonic Laplacian is given by the eigenvalue equation 
\equ{
\gD \gf + \gl \gf = 0, 
\qquad 
\gD = \der_y^2 - g_5 q_b \Bigl( \der_y \left<\gF\right> 
+ g_5 q_b \left<\gF\right>^2 \Bigr), 
}
where we concentrate on the case where $\left<\gF\right>$ is given by 
eq.\ \eqref{gFSusy}, with the appropriate boundary conditions of the 
even and odd scalar fields imposed. As usual this ensures that the
spectrum of eigenvalues $\gl$ is discrete. 
To avoid having to deal with the second derivative of the
delta-function explicitly,  we perform the substitution 
\(
\gf = \exp \left\{ g_5 q_b \int_0^y \d y\left< \gF \right>\right\} \tgf
\)
so as to obtain the boundary value problem 
\equ{
\tgf'' + 2g_5 q_b \left< \gF \right> \tgf' + \gl\, \tgf = 0, 
\qquad 
\tgf'(0) = \tgf'(\gp R) = 0,
\labl{BoundaryValue}
}
for the even fields. (For odd fields the boundary conditions are 
$\tgf(0) = \tgf(\gp R) = 0$.)

As this equation still involves a first derivative on the delta
function, some appropriate regularization is needed. In appendix 
\ref{a:ScalarSpectrum} such a regulator is provided and the algebraic
equation that determines the eigenvalues is obtained. In the limit
where this regulator is removed, the spectrum for various signs of the
FI--parameters and even and odd fields becomes identical, and
completely independent of the FI--parameter $\gx_0''$. The mass
spectrum reads 
\equ{
\gl_n = \frac 14 (g_5 q_b \gx_0)^2 + \frac {n^2}{R^2}, 
\quad 
n \in \Natr, 
\labl{Spectrum}
}
see \eqref{Spec0}. Clearly, in the limit $|\gx_0| \ra \infty$ (i.e.\
when the cut-off $\gL$ becomes extremely large) this shows that all
non-zero modes become infinitely heavy, and that all these states
should decouple from the theory. However, there might be some
subtleties due to non-decoupling effects as we will see in our
analysis of anomalies.

\section{Anomalies}
\labl{s:Anomalies}

In this section we turn to the important discussion of anomalies in
orbifold models. The first subsection is devoted to the analysis of 
a specific type of global anomaly for theories on $\cM^4 \times S^1$: 
the parity anomaly. The importance of global anomalies has been
studied extensively in the past
\cite{Alvarez-Gaume:1985dr,Witten:1985xe}.  
One of the first works on this was \cite{Witten:1982fp}, which showed 
that a theory with an odd number of $SU(2)$ doublets is inconsistent. 
We will see that the parity anomaly leads to a similar conclusion for 
MSSM-like orbifold models with an odd number of $SU(2)$ doublets 
or $SU(3)$ triplets in the bulk. The 
central observation we make in this section is that before defining an 
orbifold model one should check whether the parity symmetry is a
quantum symmetry because otherwise it may be problematic to divide it
out to construct the orbifold. 

The next section studies the local gauge anomalies that can arise in
orbifold models. Using topological arguments it is shown that the
gauge anomalies appear only on the boundaries. Conditions for
consistency of an orbifold theory with fermions is given. In the last
subsection we discuss to what extend the localization effects can
change the discussion of anomalies.

\subsection{Parity anomaly on $\boldsymbol{S^1}$}
\labl{s:parity_anom}

In this subsection we point out that some orbifold models containing
fermions are ill--defined. To explain what the problem is, we need to
take one step back and consider a fermionic field theory on a
circle in 5 dimensions coupled to a gauge field $A_M$ associated to a
Lie group $G$. 
On the circle this theory is classically invariant under the parity  
transformation
\equ{
y\to-y, \quad
A_{\gm}(y) \to A_{\gm}(-y), \quad
A_{5}(y) \to -A_{5}(-y), \quad
\gps(y) \to i \gg^5 \gps(-y). 
\labl{S1parity} 
}
In the construction of the orbifold $S^1/\Intr_2$ this 
symmetry is divided out.  

But this symmetry can be anomalous as was observed in ref.\  
\cite{Alvarez-Gaume:1985nf}, and can lead to a sign ambiguity in the 
fermionic determinant. This anomaly can be canceled by adding
the parity anomaly counter term 
\equ{
\gG_{PAC}(A) = 
- \gp i\,  \int_{\cM^{4} \!\times\! S^1} \gO_{5}(A),
\labl{parity_counter}
}
with $\gO_5(A)$ the Chern-Simons $5$-form defined in eq.\ 
\eqref{descent} of appendix \ref{a:charClasses}. (There we briefly
recall some standard facts about anomalies and characteristic classes
to set up the notation used in the section.) The effective (Euclidean)
quantum action $\gG(A)$ with the fermions integrated out is defined only 
modulo $2\gp i$, see eq.\ \eqref{pathint}. Therefore, if the difference
$\gD\gG_{PAC}(A)$ between the parity transformed and the counter 
term \eqref{parity_counter} satisfies 
\equ{
\gD\gG_{PAC}(A) = 2\gp i\,  \int_{\cM^{4} \!\times\! S^1} \gO_{5}(A) = 0\
\text{mod}\ 2\gp i, 
\labl{AbsentParAnom}
}
the model does not have a parity anomaly.
Since the  APS $\get$-invariant \cite{Atiyah:1980jh}
\(
\get = 2 \int  \gO_{5}(A) + \text{const}.
\)
is an integer, it follows that if $\int \gO_{5}(A)$ is an even
multiple of $\int \gO_{5 |F}(A)$ computed with the trace of the
fundamental representation, there is no parity anomaly. As this 
Chern-Simons form is proportional to the symmetric trace 
$D_{ijk} = \tr ( T_i \{T_j, T_k\})$, standard four 
dimensional anomaly cancelation arguments can be used to 
argue when this counter term is just zero. 

It is not hard
to show that the parity counter term is invariant under infinitesimal 
gauge transformations, using similar arguments as in 
\eqref{odd_gauge_anom}. However, this counter term $\gG_{PAC}$ 
gives back a sign ambiguity by a large gauge transformation $g$ 
if 
\equ{
\gG_{PAC}({}^g A)  \neq 0\ \text{mod}\ 2\gp i.
\labl{ParCountGauge}
}
If this happens the counter term $\gG_{PAC}$ cannot be added 
since then gauge invariance is lost.  In appendix 
\ref{a:gauge_parity} we analyse, using homotopy groups, which 
parity anomaly counter terms are potentially not gauge invariant. 
We find that if the gauge group contains $SU(n)$ (with $n \geq 3$) or 
$USp(2n)$ factors, or is a product of at least one $U(1)$ and 
some simple groups, gauge invariance of the parity counter term
depends on the precise choice of matter representation, otherwise
gauge invariance is automatic. 
(Parity counter term for just $U(1)$ can also violate gauge symmetry
if gravitational interactions are also taken into account.)
 
The parity anomaly for a field theory on the circle $S^1$ is not
a problem, it just means that the parity invariance is broken
at the quantum level. However, if one wants to define a field theory
on the orbifold $S^1/\Intr_2$, it is of course crucial that this
parity symmetry is an exact symmetry of the quantum theory, else 
it makes no sense to mod it out!

Therefore, if the fermionic determinant on $S^1$ is not parity
invariant, the parity counter term \eqref{parity_counter} has to be
added. But then it is crucial that gauge invariance in the sense
of eq.\ \eqref{ParCountGauge} is not lost! This reintroduces a
sign-ambiguity in the path integral description of the theory with
fermions and gauge fields. It is important to note
here that the orbifolding condition does not project out those 
large $U(1)$--gauge transformations under which the parity counter term
is not invariant. For example, they can be maps of $g: S^1 \ra U(1)$ that are 
topologically equivalent to the form $g(y) = e^{i y/(2\pi R)}$ which 
has non-trivial winding on $U(1)$. Hence the parity changes the
orientation of the winding but does not undo it. 
If it is not possible to
save the parity symmetry on $S^1$ using a gauge invariant counter
term, then one can question whether the model with the parity divided
out at the classical level, gives rise to a consistent quantum field theory. 

We now investigate which MSSM-like models with fields in the bulk
and/or on the branes do not suffer from the parity anomaly. According
to the analysis above, we have to check that all possible parity
anomaly counter terms appear as even multiples of those for the
fundamental representations. This gives the following constraints for
the bulk multiplets: 
\equ{
\left. 
\arry{c l}{
U(1)^3 & 
6 n_Qq_Q^3 + 2 n_Lq_L^3 + 3 n_{u^*}q_{u^*}^3 + 3
n_{d^*}q_{d^*}^3 + n_{e^*}q_{e^*}^3 + 2 n_{H_1}q_{H_1}^3 + 2 n_{H_2}q_{H_2}^3
 \\[2ex]
SU(3)^3 & 2n_Q + n_{u^*} + n_{d^*}
\\[2ex]
U(1) \times SU(2)^2 &3 n_Qq_Q + n_Lq_L + n_{H_1}q_{H_1} + n_{H_2}q_{H_2}
\\[2ex]
U(1)\times SU(3)^2 & 2 n_Qq_Q + n_{u^*}q_{u^*} + n_{d^*}q_{d^*}
 \\[2ex]
U(1)\times \text{grav} & 6 n_Qq_Q + 2 n_Lq_L + 3 n_{u^*}q_{u^*} + 3
n_{d^*}q_{d^*} + n_{e^*}q_{e^*} + 2 n_{H_1}q_{H_1} + 2 n_{H_2}q_{H_2}
}
\right\} = 0\ \text{mod}\ 2, 
}
where $q_X$ is the hypercharge of multiplet $X$ in units of the
hypercharge of the quark doublet Q and $n_X$ is the number 
of such multiplets present in the bulk. Three of the above conditions
are fulfilled automatically for arbitrary $n_X$ because $q_{u^*}$,
$q_{d^*}$ and $q_{e^*}$ are even. 
Only the second and third relations give the non trivial restrictions:
\equ{
\arry{rcl}{
n_{u^*} + n_{d^*} & = & 0\ \text{mod}\ 2\,,
\\[2ex]
n_Q + n_L + n_{H_1} + n_{H_2} & = & 0\ \text{mod}\ 2\,.
}}
So we conclude that in MSSM--like models the parity anomaly appears
when an odd number of $SU(2)$ doublets (quark, lepton or Higgs) 
or an odd number of right-handed quarks live in the bulk.
Hence we need a parity counter term that involves $U(1)$ and 
$SU(2)$, and $SU(3)$ gauge field interactions. However, this is
precisely of the  
form of not necessarily gauge invariant parity counter terms as 
explained in appendix \ref{a:gauge_parity}. Indeed for an odd number
of $SU(2)$ doublets or an odd number of $SU(3)$ triplets of the MSSM the
conclusion is that the parity counter term is not gauge invariant. 

Let us close this subsection by making some general comments 
clarifying the meaning and further applications of the results
presented here.  First of all an orbifold theory is a special theory
on an interval in which, apart from 
the Dirichlet or Neumann boundary conditions for the fields, there is
a symmetry that forbids many operators. By doubling the space this
symmetry can be made explicit and is identified with the $\Intr_2$
parity. Therefore the parity anomaly does not forbid a quantum field
theory on an interval with appropriate boundary conditions, but it
implies that many additional terms may appear in the Lagrangian as
there is no (quantum) symmetry which forbids them. 

Secondly, we notice that the parity anomaly may have (important) 
consequences when the mod out of the parity symmetry is
used to break the gauge symmetry \cite{Kawamura:2000nj} (which may
have interesting applications like solving the doublet-triplet
splitting problem \cite{Kawamura:2001ev}). Here it is important to
take into account the distinction between inner and outer automorphism
gauge symmetry breaking (see e.g.\ \cite{Hebecker:2002jb}). 
For inner automorphisms we see, that the parity anomaly can arise under
the same conditions as discussed in the subsection and appendix
\ref{a:gauge_parity} 
(since with $A_\mu(-y) = P A_{\mu}(y) P\inv$, etc., the gauge group
element $P$ that squares to unity, drops out of the Chern-Simons term 
\eqref{AbsentParAnom}). The parity anomaly may not arise when outer
automorphisms are used to break the gauge group: for example, under
complex conjugation of the generators of the gauge group 
$T_a \ra (-T_a)^T$, the parity counter term is invariant.

\subsection{Anomaly free models on $\boldsymbol{S^1/\Intr_2}$}

In the recent literature 
\cite{Scrucca:2002eb,Arkani-Hamed:2001is,Pilo:2002hu,Barbieri:2002ic,Kim:2002ab}
there has been a lot of attention to the
discussion on the possible forms of local gauge anomalies on orbifolds. 
On the orbifold $S^1/\Intr_2$, the standard argument that there cannot
be anomalies as there are no chiral fermions does not hold because on
the four dimensional boundaries chiral projection may appear.  In this
section we start with the possible gauge 
anomaly due to a five dimensional Dirac fermion coupled to a gauge
field $A_M$ on the orbifold $S^1/\Intr_2$. After that we consider 
localization effects, additional chiral fermions on the boundaries,
and discuss possible Chern-Simons counter terms. 

Following Horava-Witten \cite{Horava:1996qa} the form of the 
anomaly for Dirac fermion $\gps$ on $\cM_5 = \cM^4 \times S^1/\Intr_2$, 
that satisfies $\gps(-y) = i \gg^5 \gps(y)$, can be determined as
follows. First of all the anomaly has to be mathematically consistent,
this means that the anomaly has to satisfy the Wess-Zumino 
consistency condition \cite{Wess:1971yu}. Since the solution
$\gO_4^1(A, \gL)$ to this condition is unique up to a normalization
factor,  the effective action $\gG(A)$, obtained by integrating the
fermion out, transforms as  
\equ{
\gd_\gL \gG(A)
= \gp i \sum_{I = 0, \gp} \tilde n_I \left.  \int_{\cM^{4}} \gO^1_{4}(A; \gL) 
\right|_{I R},
\qquad 
e^{-\gG(A)} =  \int \cD \gps \cD \bgps \, e^{- S },
\labl{orbi_gauge_anom}
}
with only the constants $\tilde n_I$ left to be determined. (One could have
added  
\(
\int_{\cM_5} d \gO^1_4(A,\gL), 
\)
but this can be rewritten as two boundary terms, already considered 
in \eqref{orbi_gauge_anom}.) This argument shows that the anomaly can
only appear on the 4 dimensional boundaries. Since the projections on
both fixed points are the same, it implies that $\tilde n_0 = \tilde n_\gp$. By
restricting both the gauge parameter and the fields to four
dimensions, the standard result for a chiral fermion in 4 dimensions
should be obtained: eq.\ \eqref{gauge_anom} of the appendix
\ref{a:charClasses}. This then fixes the normalization completely 
$\tilde n_0 = \tilde n_\gp =1$. 

Writing the anomaly as an integral over the five dimensional space
$\cM_5$ 
\equ{
\gd_\gL \gG(A) = 
\gp i \int_{\cM_5} \left( \gd(y) + \gd(y - \gp R) \right) 
\gO^1_{4}(A; \gL) \d y,
}
the same form of the anomaly is obtained as the one  
found by \cite{Arkani-Hamed:2001is} using a perturbative
calculation. It was observed there that the result was independent
of the shape of the modes used to perform the calculation, provided of
course, that the set of modes is complete. This conclusion is also
reached from the reasoning given here: we see clearly the
topological origin of this anomaly as it is defined in terms of the
characteristic classes.   

Next, we include the possibility of having chiral fermions on the
boundaries. Under a gauge transformation the effective actions
$\gG_I(A)$ of the fermions on the branes $I= 0, \gp$ can have
anomalies 
\equ{
\gd_\gL \gG_I(A)
=  2\gp i \left.  \int_{\cM^{4}} \gO^1_{4 |I}(A; \gL) 
\right|_{I R},
\qquad 
e^{-\gG_I(A)} =  \int \cD \gps_I \cD \bgps_I \, e^{-S_I },
\labl{brane_gauge_anom}
}
with $|I$ we indicate that the trace which is implicit in the form
$\gO^1_{4|I}$ is taken over the chiral fermions living on brane $I$
only.

In order to obtain a consistent model, we may need to
introduce possible Chern-Simons terms. The uniqueness 
of the descent equation \eqref{descent} tells us 
that there is only one possible five dimensional Chern-Simons action
$\gG_{CS}(A)$ available up to normalization 
\equ{
\gG_{CS}(A)  = N_{CS}\, \gp i \int_{\cM_5} \gO_{5 |F}(A),
}
where the internal trace is taken over the fundamental
representation ($F$) to fix the normalization $N_{CS}$ uniquely. 
The gauge transformation \eqref{charCl} of the Chern-Simons term reads 
\equ{
\gd_{\gL} \gG_{CS}(A) = 
N_{CS}\, \gp i  \int_{\cM_5} d \gO^1_{4|F}(A; \gL) 
= N_{CS}\, \gp i \int_{\cM_5}\left( -\gd(y) + \gd(y - \gp R) \right) 
\gO^1_{4 |F}(A; \gL) \d y.
\labl{CS_anom}
}
The minus sign in front of $\gd(y)$ appears because of changing the
induced orientation, due to Stoke's theorem, to standard one.

A model in five dimensions on the orbifold $S^1/\Intr_2$ with both
bulk and brane fermions can be made locally consistent, if one can find a
normalization $N_{CS}$ of the Chern-Simons term, such that the total
effective action  
\equ{
\gG_{tot}(A) = \gG(A) + \gG_0(A) + \gG_\gp(A) + \gG_{CS}(A)
\labl{total_eff_action}
}
is gauge invariant under local (in general five dimensional) gauge
transformations. By normalizing the bulk and brane fermion
anomalies \eqref{orbi_gauge_anom} and \eqref{brane_gauge_anom} 
w.r.t.\ the fundament representation ($F$), 
\equ{
\arry{lcl}{
\gd_{\gL} \gG(A) 
&=&\dsp ~~ 
N\, \gp i \int_{\cM_5}\left( \gd(y) + \gd(y - \gp R) \right) 
\gO^1_{4 |F}(A; \gL) \d y, 
\\[2ex] 
\gd_{\gL} \gG_{I}(A) 
&= & \dsp 2 N_{I}\, \gp i \int_{\cM_5} \gd(y - I R)  
\gO^1_{4 |F}(A; \gL) \d y,
}
}
similarly to the Chern-Simons term, one can determine $N_{CS}$,
and obtain a consistency condition
\equ{
N_{CS} = N_0 - N_\gp,
\qquad 
N + N_0 + N_\gp = 0. 
}
Notice that the consistency requirement takes the form of a sum rule
and is determined by the fermionic content of the bulk and branes
only. Furthermore, we see that a Chern-Simons term is required only if
the anomalies due to both branes are not equal. It should be
stressed that the results of this subsection depend on properties of
the fermionic content of the model, therefore the requirements should
hold in both supersymmetric and non-supersymmetric models on
$S^1/\Intr_2$.\footnote{
A similar analysis can be preformed for the orbifold 
$S^1/\Intr_2\times \Intr_2'$. In this case the chiral projection on
one brane can be opposite to that on the other brane. Then one
finds $\tilde n_0 = - \tilde n_\gp = 1$ in \eqref{orbi_gauge_anom}. By
denoting the 
normalization of the anomaly in both cases by $N_\pm$, and 
following the same arguments, we find a consistency requirement: 
$N_+ + N_0 + N_\gp = 0$, involving matter with chiral zero modes only,
while the normalization of the Chern-Simons term is given by 
$N_{CS} = N_- + N_0 - N_\gp$.
}

It is easy
to apply these formulae to the case of possible pure $U(1)$ gauge and
mixed $U(1)$-gravitational anomalies, we find the anomaly cancelation
conditions 
\equ{
\tr (q^3) + \tr (q_0^3) + \tr (q_\gp^3) =0, 
\qquad
\tr (q) + \tr(q_0) + \tr(q_\gp) = 0. 
}
Observe that these relations are exactly the expected ones by looking
only at the low-energy spectrum determined by the zero modes. 
And the coefficients for the pure $U(1)$ and mixed Chern-Simons
counter terms are given by 
\equ{
N_{CS}(U^3(1)) \sim\tr(q_0^3) - \tr(q_\gp^3),
\qquad 
N_{CS}(\text{mixed}) \sim \tr(q_0) - \tr(q_\gp),
}
up to normalization to the minimal charge.

\subsection{Gauge anomalies on $\boldsymbol{S^1/\Intr_2}$ and
localization} 
\labl{s:AnomLocal}

The analysis of the previous subsection was performed for a quite
general 5 dimensional model on the interval $S^1/\Intr_2$ with fermions and gauge
fields. We now return to a much more restrictive situation of a
supersymmetric model.  

Because the result for the anomaly is independent of which complete set
of modes is used \cite{Arkani-Hamed:2001is}
(but of course with the same boundary conditions), 
it follows that the localization effects can not
change the form of the anomaly. However, this does not
mean that the localization does not change our description of the
situation involving anomalies. When localization of
the zero modes happen and the cut-off is taken to be very large,  
all modes except the zero mode get infinitely heavy, see section 
\ref{s:MassSpec}. Therefore for many practical purposes they
have completely decoupled from the theory. Let us introduce some 
notation to describe the situation: Let $\gG^0(A)$ denote the
effective action after integrating out the (localized) chiral zero
mode (while neglecting the massive modes) and
$\gG^M(A)$ the effective action obtained by integrating out the
massive fermionic states (and dropping the massless one). 
Since the (localized) chiral zero mode
and the massive fermion modes are independent in the path integral, it
follows that the bulk effective action may be written as 
$\gG(A) = \gG^0(A) + \gG^M(A)$. From this equation, one can define
the ``massive'' anomaly due to the (infinitely) heavy modes by
\equ{
\gd_\gL \gG^M(A) = \gd_\gL\gG(A) - \gd_\gL\gG^0(A).
}
As the first term is given by \eqref{orbi_gauge_anom}, and the second
we know since it represents the anomaly due to a
localized state, the ``massive'' anomaly is given. As an 
example, let us give a graphical representation of this equation. 
Consider the case that the zero modes get localized at brane $0$, this
leads to the identity 
\equ{
\raisebox{-13mm}{
{\scalebox{.8}{\mbox{\input{mass_anom.pstex_t}}}}
}
{\ =\ } 
\raisebox{-13mm}{
{\scalebox{.8}{\mbox{\input{bulk_anom.pstex_t}}}} 
}
{\ -\ } 
\raisebox{-13mm}{
{\scalebox{.8}{\mbox{\input{local0_anom.pstex_t}}}}
}
{.} 
\labl{MassiveAnomaly}
}
This ``massive'' anomaly has the same structure as the 
 graphical representation of the gauge
transformation of the Chern-Simons term \eqref{CS_anom}.
In a consistent model, one would expect that this ``massive'' anomaly
is canceled by other high-energy physics effects like five dimensional 
Chern-Simons terms (see e.g.\ \cite{Kim:2002ab}), like it is done here
when the zero mode splits up to exist on both branes only.

Let us consider a simple situation, that we have considered in
our previous work  
\cite{GrootNibbelink:2002wv}: a model with one chiral
multiplet on brane $0$ and a hyper multiplet of the opposite charge in
the bulk. 
It is not hard to check that this model is anomaly free but
needs ($N_{CS} = -1$) a $U(1)$ gauge Chern-Simons term (neglecting
gravitational interaction here for simplicity) because the anomalies
on both branes are not equal. Pictorially the anomaly cancelation
in this model can be drawn as 
\equ{
\raisebox{-14mm}{
\scalebox{.8}{\mbox{\input{brane_anom.pstex_t}}}
}
{\ +\ } 
\raisebox{-13mm}{
\scalebox{.8}{\mbox{\input{bulk_anom.pstex_t}}}
} 
{\ -\ } 
\raisebox{-13mm}{
\scalebox{.8}{\mbox{\input{CS.pstex_t}}}}
\ = \ 0.
} 
Applying our analysis of section \ref{s:localization} to this model,
we infer the bulk field get localized at brane $0$. Hence we may use
the equality \eqref{MassiveAnomaly} to eliminate the gauge variation
of the Chern-Simons term. The bulk anomaly contribution cancels
out and we are left with local anomaly cancelation on the $0$ brane. 
\equ{
\raisebox{-14mm}{
\scalebox{.8}{\mbox{\input{brane_anom.pstex_t}}}
}
{\ +\ } 
\raisebox{-13mm}{
\scalebox{.8}{\mbox{\input{local0_anom.pstex_t}}}
} 
\ = \ 0.
} 
After the localization the Chern-Simons term is absent, canceled
so to say, against the anomaly due to the heavy modes, and the anomaly
cancelation involves zero modes only.

\section{Conclusions}

In this work we have set out to investigate two aspects of
supersymmetric theory in 5 dimensions compactified on the orbifold
$S^1/\Intr_2$ which may lead to a deeper understanding of structure
and properties of such models: consistency requirements due to
anomalies and instabilities due to Fayet-Iliopoulos terms. 

After a review of supersymmetric theories in 5 dimensions with
boundaries, we performed a one-loop calculation of the
Fayet-Iliopoulos terms at the branes generated by brane and bulk
fields. We showed that the bulk fermions are responsible for the
generation of the tadpole for $\der_y \gF$, which is required to
appear together with the auxiliary field $-D_3$ because of the
unbroken supersymmetry. 

A substantial part of the paper was devoted to the study of
localization 
effects caused by (quadratically and logarithmically) divergent
Fayet-Iliopoulos terms due to bulk and brane fields. We have studied
the localization of zero modes in the presence 
of supersymmetric and $U(1)$ gauge symmetry preserving vacuum
solutions. We have identified three different types of possible
localized zero modes when the cut-off is taken very large:
localization at a brane, localization infinitely 
close to (though not on) a brane, and simultaneous localization at
both boundaries. The second form of localization leads to the 
situation that some fields live at the same place but any direct
interaction between them is forbidden. Third
type of localization leads to states that live and interact on two
branes which may have arbitrary large distance between them. 

It should be stressed that the localization discussed in this paper is 
typically of the delta-like type. The reason for this is the
$\gd''$-contribution in the FI--tadpoles. 
Only the second kind of localization
(close to a brane) may have some finite width which is controlled by
the cut-off scale $\gL$ (so this width may be substantial only when
the cut-off scale is very low or when the tree level FI--term is
fine--tuned against the radiatively generated one). This means that
the localization effect is important, not only for models with
constant 
zero modes of bulk fields, but also for models (for example 
\cite{Kaplan:2001ga,Arkani-Hamed:2000dc,Georgi:2001wb})
in which the zero modes have more complicated shapes.

Another source of quantum instabilities, anomalies, were investigated
in the remainder of the paper. First, we discussed a possible
difficulty in defining the fermionic model on this orbifold at the
quantum level, if on $S^1$ there is an anomaly in the $\Intr_2$ parity
symmetry. We derived sufficient requirements on the fermionic spectrum
of theory, so as to avoid these complications. 

Then we moved to local gauge anomalies on the orbifold
$S^1/\Intr_2$. Using the uniqueness of the solutions of the
Wess-Zumino consistency condition for anomalies, the structure of 
anomalies due to bulk and brane fields could be easily determined. 
The role of possible Chern-Simons terms in anomaly cancellation was
addressed: we showed that demanding 5 dimensional local gauge
invariance leads to consistency conditions which are independent of 
the presences of Chern-Simons interactions. These conditions are
nothing but the familiar 4 dimensional anomaly cancellation
requirements for the zero modes of the theory. 
The normalization of the Chern-Simons terms, on
the other hand, is determined by the imbalance of gauge anomalies due
to the chiral fermions on the two branes. In addition, we showed, that
although the localization of bulk zero modes to branes does not change
the consistency requirement on the theory (since this only involve 
the zero modes), the remaining (infinitely) heavy modes may cancel
(parts of) the Chern-Simons terms.

\section*{Acknowledgments}

We would like to thank H.M. Lee, J.E. Kim and M. Walter for useful
discussions. 
\\ 
Work supported in part by the European Community's Human Potential
Programme under contracts HPRN--CT--2000--00131 Quantum Spacetime,
HPRN--CT--2000--00148 Physics Across the Present Energy Frontier
and HPRN--CT--2000--00152 Supersymmetry and the Early Universe.
SGN was supported by priority grant 1096 of the Deutsche
Forschungsgemeinschaft. 
MO was partially supported by the Polish KBN grant 2 P03B 052 16.

\appendix
\def\theequation{\thesection.\arabic{equation}} 

\section{Notations and Conventions}
\labl{a:conventions}
\setcounter{equation}{0}

The metric has the signature $\get_{MN} = (+1,-1,-1,-1,-1)$,
with $M, N = 0, \ldots 4$. The Clifford algebra in five dimensions can
be characterized by the following relations
\equ{
\{ \gg^M, \gg^N \} = 2 \get^{MN},
\quad
{\gg^M}^\dag = \gg^0 \gg^M \gg^0,
\quad
C_+\inv \gg^M C_+ = {\gg^M}^T,
\quad
C_+^T = - C_+,
\quad
C_+^\dag = C_+\inv.
}
Dirac and charge conjugation are given by
\equ{
\bgps = \gps^\dag \gg^0,
\qquad
\gps^{C_+} = C_+ \bgps^T,
}
respectively.

The symplectic structure of $N=1$ in five dimensions can be made
manifest using symplectic (Majorana) reality conditions
\equ{
\gve = \ge \gve^{C_+},
\quad
\gch = \ge \gch^{C_+},
\quad
\gz = \gr \gz^{C_+},
\quad
h = \gr h^* \ge\inv,
}
for the supersymmetry parameters $\gve$, the gauginos 
$\gch = (\gch^\ga)$, the hyperinos $\gz = (\gz^a)$ and the hyperons
$h = (h^{a\ga})$, with $\ga = 1,2$ and $a = 1,\ldots 2n$ for $n$ hyper
multiplets, respectively.
The chirality operator $i\gg^5$ and the matrices $\ge$ and $\gr$ can
be represented as
\equ{
i \gg^5 = \pmtrx{\Id & ~~0 \\ 0 & - \Id},
\qquad
\ge = \pmtrx{ ~~0 & 1 \\ -1 & 0},
\qquad
\gr = \Id \otimes \ge = \pmtrx{ ~~0 & \Id_n \\ -\Id_n & 0}.
}
The five and four dimensional charge conjugation matrices are related
via $C_+ = i \gg^5 C_-$.

The Pauli-matrices, denoted by $\vec \gs = (\gs^A)$, carry the same
indices as the auxiliary fields $D^A$ of the vector multiplet. The
different indices are suppressed as much as possible throughout this
work.

\section{Scalar spectrum} 
\labl{a:ScalarSpectrum}
\setcounter{equation}{0}

In this appendix we compute the spectrum of the boundary value 
problem \eqref{BoundaryValue} for even fields in detail. Since the 
calculation for the odd fields is completely analogous, we conclude
this appendix by simply quoting the results in that case. 

The spectrum of the bosonic Laplacian is given by the eigenvalue
equation  
\equ{
\gD \gf + \gl \gf = 0, 
\qquad 
\gD = \der_y^2 - g_5 q_b \Bigl( \der_y \left<\gF\right> 
+ g_5 q_b \left<\gF\right>^2 \Bigr), 
}
where we concentrate on the case where $\left<\gF\right>$ is given by 
eq.\ \eqref{gFSusy}. Taking $\gf$ to be even implies that on the
interval $[0, \gp R]$ it has to satisfy the boundary conditions 
$\gf'(0) = \gf'(\gp R) = 0$. As usual this ensures that the spectrum
of eigenvalues $\gl$ is discrete. 

To avoid having to deal with the square of the first derivative of 
delta--functions explicitly,  we perform the substitution 
\(
\gf = \exp \left\{ g_5 q_b \int_0^y \d y\left< \gF \right>\right\} \tgf
\)
so as to obtain the boundary value problem 
\equ{
\tgf'' + 2g_5 q_b \left< \gF \right> \tgf' + \gl\, \tgf = 0, 
\qquad 
\tgf'(0) = \tgf'(\gp R) = 0.
\labl{boundtgf}
}

As this equation still contains first derivatives of delta functions,
we take the regularized form of the delta function and its derivative 
to be
\equ{
\gd_\gr(y) = 
\begin{cases}
{\dsp 
\frac 1\gr \left( 1 - \frac {|y|}{\gr} \right)} & |y| <\gr, 
\\[2ex]
0 & |y| > \gr,
\end{cases}
\qquad
\gd_\gr'(y) = 
\begin{cases}
- {\dsp \frac 1{\gr^2}\, \sgn(y)} & |y| < \gr, 
\\[2ex]
0 & |y| > \gr.
\end{cases}
}
Employing this regularization means that we have to solve three 
differential equations with the appropriate (continuously
differentiable) matching and boundary conditions. By introducing
somewhat more elaborate notation the three cases can be treated
simultaneously. 
\equ{
\tgf'' + 2 \gm_{(a)} \tgf' + \gl\, \tgf = 0,
\qquad 
\gm_{(a)} = g_5 q_b \left( \half \gx_0 - a \, \frac{\gx_0''}{\gr^2} \right),
\labl{DEtgf}
}
where $a = +, 0, -$ refers to the regions $[0, \gr[$, 
$]\gr, \gp R- \gr[$ and $] \gp R- \gr, \gp R]$, respectively. 
The solutions of the equation are, of course, exponentials 
\(
\tgf(y) = \exp (\ga_{(a)}^b y )
\)
with the exponents 
\equ{
\ga_{(a)}^b = - \gm_{(a)} + b 
\sqrt{\gm_{(a)}^2 - \gl},
\quad b = \pm, 
\qquad 
\ga_{(a)}^+\ga_{(a)}^- = \gl.
\labl{DEexp}
}
For $\ga_{(0)}^\pm$ we often use the short hand $\ga^\pm$. Notice that
if $\gx_0'' = 0$, for all $a$ we have that $\ga_{(a)}^b = \ga^b$. 
The solution, with the boundary conditions \eqref{boundtgf} taken into
account, can be represented as  
\equ{
\tgf(y) = 
\begin{cases}
A_{(+)} \left(
 \ga_{(+)}^- e^{\ga_{(+)}^+ y} - \ga_{(+)}^+ e^{\ga_{(+)}^- y}
\right), & 0 < y < \gr, 
\\[2ex]
A^+ e^{\ga^+ y } + A^- e^{\ga^- y}, & 
\gr < y < \gp R - \gr,
\\[2ex]
A_{(-)} \left(
 \ga_{(-)}^- e^{\ga_{(-)}^+ (y-\gp R)} 
- \ga_{(-)}^+ e^{\ga_{(-)}^- (y-\gp R)}
\right), & \gp R - \gr < y < \gp R.
\end{cases}
}
By gluing the solutions together at the two boundaries between the
three regions in a continuously differentiable fashion,  allows one to
express $A^\pm$ linearly in terms of either $A_{(+)}$ or $A_{(-)}$. 
Therefore, in the ratio $- A^+/A^-$ these constants drop out and a
consistency condition resulting from the boundary conditions is
obtained. This consistency requirement 
\equ{
e^{(\ga^+ - \ga^-)(\gp R - 2 \gr)} = 
\frac {1 - \ga^- H_{(-)}}{1 - \ga^+ H_{(-)}} \, 
\frac {1 - \ga^+ H_{(+)}}{1 - \ga^- H_{(+)}}, 
\qquad 
H_{(a)} = \frac 1{\ga_{(a)}^+ \ga_{(a)}^-}
\frac{
 \ga_{(a)}^- e^{a \ga_{(a)}^+ \gr} - \ga_{(a)}^+ e^{a \ga_{(a)}^- \gr}
}{e^{a \ga_{(a)}^+ \gr} - e^{a \ga_{(a)}^- \gr}}, 
\labl{Specgr}
}
leads to the quantization of the eigenvalue $\gl$. 

As a consistency check of this equation we observe that in the case 
$\gx_0'' = 0$, dividing the interval $[0, \gp R]$ into three regions is 
not necessary and therefore the result of the spectrum should be
independent of $\gr$. Indeed, in this case we find that the 
quantization condition \eqref{Specgr} reduces to 
\equ{
e^{(\ga^+ - \ga^-)\gp R } = 1 
\quad \Ra \quad 
\gl_n = \frac 14 (g_5 q_b \gx_0)^2 + \frac {n^2}{R^2}, 
\quad 
n \in \Natr, 
\labl{Spec0}
}
for any $\gr$. After the implication sign we used eqs.\ \eqref{DEtgf}
and \eqref{DEexp} and that eq.\ \eqref{Spec0} has solutions only 
if  $\gl> \frac 14 (g_5 q_b \gx_0)^2$. (For equality, the general
solution is not of the form of a sum of two exponentials, and does not
allow for non-trivial solutions of the boundary conditions. This
excludes the would be zero mode $n =0$, which we have already treated
in detail in section \ref{s:shapeZero}.) 

For $\gx_0'' \neq 0$, we are interested in the limit $\gr \ra 0$ where
the delta function and its derivative is reproduced. Although the
asymptotics 
\equ{
H_{(\pm)} \ra 
\begin{cases}
\pm 0 & \gx_0'' q_b > 0, 
\\[2ex]
\mp \infty & \gx_0'' q_b <0, 
\end{cases}
\labl{asympH}
}
of the functions $H_{(\pm)}$ is very different for the cases 
$\sgn(\gx_0'' q_b)  = \pm$, in the limit of 
$\gr \ra 0$ of \eqref{Specgr} leads to the same quantization condition 
\eqref{Spec0}. (Here $\pm 0$ denotes where the limit $0$ is reached
from above or below.) With this result we have shown, that it does not
matter for the spectrum whether the term with $\gx_0''$ is present or not,
provided that the branes are infinitely thin ($\gr \ra 0$); as is
dictated by the orbifolding. 

For the odd bosonic states, we can perform a completely analogous
analysis, therefore we just quote the main results. The boundary
conditions for odd fields $\tgf(0) = \tgf(\gp R) =0$, is implemented
by the solution 
\equ{
\tgf(y) = 
\begin{cases}
A_{(+)} \left(
 e^{\ga_{(+)}^+ y} -e^{\ga_{(+)}^- y}
\right), & 0 < y < \gr, 
\\[2ex]
A^+ e^{\ga^+ y } + A^- e^{\ga^- y}, & 
\gr < y < \gp R - \gr,
\\[2ex]
A_{(-)} \left(
e^{\ga_{(-)}^+ (y-\gp R)} - e^{\ga_{(-)}^- (y-\gp R)}
\right), & \gp R - \gr < y < \gp R.
\end{cases}
}
This gives rise to the quantization equation of the form as in the even
case, see \eqref{Specgr}, except the functions $H_{(\pm)}$ are
different: 
\equ{
e^{(\ga^+ - \ga^-)(\gp R - 2 \gr)} = 
\frac {1 - \ga^- H_{(-)}}{1 - \ga^+ H_{(-)}} \, 
\frac {1 - \ga^+ H_{(+)}}{1 - \ga^- H_{(+)}}, 
\qquad 
H_{(a)} = 
\frac
{e^{a \ga_{(a)}^+ \gr} - e^{a \ga_{(a)}^- \gr}}{ 
\ga_{(a)}^- e^{a \ga_{(a)}^+ \gr} - \ga_{(a)}^+ e^{a \ga_{(a)}^-
\gr} }.
\labl{SpecgrOdd}
}
However the asymptotics is identical to \eqref{asympH}, so that the same
conclusion is reached: the spectrum is independent of $\gx_0''$ as long
as $\gr \ra 0$, and is given by \eqref{Spec0}.

\section{Characteristic classes and anomalies}
\labl{a:charClasses}
\setcounter{equation}{0}

It is well--known (see for example \cite{Nakahara:1990th,Alvarez2}) 
that the form of possible counter terms for anomalies
is dictated by characteristic classes. For chiral (non-Abelian)
anomalies in $2n$ dimensions, the starting point is the formal
$2(n+1)$ form
\equa{
\gO_{2(n+1)}(R, F) & =  \left. \hat {\text{A}}(R) \text{ch}(F)
\right|_{2n +2}, 
\labl{charCl}
 \\[1ex]
\hat {\text{A}}(R) & = \tr \frac {R/4\gp} {\sinh (R/4\gp)},
\quad
\text{ch}(F) = \tr \exp \frac {i F}{2 \gp}.
\non
}
Here $\text{ch}(F)$ denotes the Chern
character of the field strength 2-form $F$ of a gauge connection $A_M$;
and $\hat{\text{A}}(R)$ the roof genus of the curvature 2-form $R$. 
In this work we focus primarily on gauge anomalies, neglecting the
gravitational and mixed anomalies, therefore we do not take the roof genus
contribution into account from this point onwards (except when we
comment on mixed gauge-gravitational-like anomaly contributions). 

By the descent equations the following $2n\!+\! 1$ form
$\gO_{2n+1}(A)$ and $2n$ form $\gO^1_{2n}(A; \gL)$ are defined
\equ{
\d \gO_{2n+1}(A) = \gO_{2(n+1)}(F), 
\qquad 
\gd_{\gL} \gO_{2n+1}(A) = \d \gO^1_{2n}(A; \gL),
\labl{descent}
}
with $\gL$ the infinitesimal parameter of a gauge transformation 
$\gd_\gL A = d \gL + [\gL, A]$. 

The effective (Euclidean) action is obtained by integrating out the
fermions $\gps$ 
\equ{
e^{- \gG(A)} = \int \cD \gps \cD \bgps \, 
e^{- S },
\quad 
S = - \int \d x \, \bgps D\Slashed(A) \gps. 
\labl{pathint}
}
In the covariant
derivative $D_M = \der_M + A_M$ the gauge field $A_M$ resides.  
There is no local gauge anomaly if for any local gauge parameter $\gL(x)$  
\equ{
\gd_{\gL} \gG(A) = \gG(A + \gd_\gL A) - \gG(A) = 0;
}
otherwise there is a gauge anomaly. For Minkowski-spaces gauge
anomalies can only occur in even $2n$ dimensions. The general form of
this anomaly is determined by the Wess-Zumino consistency conditions
\cite{Wess:1971yu} for anomalies and the connection with the Atiyah-Singer
index \cite{Atiyah:1968ih} to be 
\equ{
\gd_{\gL} \gG(A)
= 2 \gp i  \int_{S^{2n}} \gO^1_{2n}(A; \gL).
\labl{gauge_anom}
}
Here $2n$ dimensional Minkowski space has been replaced by its
Euclidean analog and a point at infinity is added
to obtain the topology of a $2n$ dimensional sphere $S^{2n}$. 

In odd ($2n+1$) dimensional Minkowski spaces no gauge anomalies can
appear. The reason for this is simply that integrals over closed
surfaces of closed forms are zero. The Wess-Zumino 
consistency conditions imply that the only possible form of the anomaly 
odd dimensions is 
\equ{
\gd_\gL \gG(A) = 2 \gp  i \int_{S^{2n+1}} 
\d \gO_{2n}^1(A; \gL). 
\labl{odd_gauge_anom}
}
But this vanishes identically, as a closed form is integrated over a
closed surface.

\section{Gauge invariance of the parity anomaly counter term}
\labl{a:gauge_parity}
\setcounter{equation}{0}

In this appendix we show, using topological methods, that the parity
anomaly counter term $\gG_{PAC}$ is not necessarily gauge invariant
\cite{Alvarez-Gaume:1985nf}, if the gauge group $G$ contains 
a factor of $SU(n)$ (with $n \geq 3$) or $USp(2n)$, or at least one
$U(1)$ factor and a simple compact Lie group.

Let us first assume that the group $G$ is connected. Under a large
gauge transformation the parity counter term variation equals the
winding number of the gauge transformation in the group
\equ{
 \gG_{PAC}({}^g A) = 
- \gp i \,  \frac {2}{5!} \Bigr( \frac i{2\gp} \Bigr)^{3}
\int_{S^4 \times S^1} \tr (g\inv \d g)^{5},
}
which can be non-zero if the gauge transformation is a non-trivial map
from $S^4 \times S^1$ or $S^5$ into the gauge group $G$.\footnote{
$S^5$ appears when we consider gauge transformations satisfying the
condition $g(x, 0)=g(x, 2 \gp R)=g$ for some constant $g$. In such a
case $S^1 \times S^4$ becomes topologically equivalent to $S^5$}
Therefore there are two possibilities: either both $\gp_4(G)$ and
$\gp_1(G)$, or $\gp_5(G)$ have to be non-trivial.  
According to the table in \cite{Weinberg:1996kr} 
\equ{
\gp_1(G) = 
\begin{cases}
\Intr & U(1), \\
\Intr_2 & SO(n \geq 3), \\
0 & \text{others},
\end{cases}
\quad 
\gp_4(G) = 
\begin{cases}
\Intr_2 \times \Intr_2 & SO(4), Spin(4), \\
\Intr_2 & SU(2), SO(3), Spin(5), SO(5), USp(2n), \\
0 & \text{others}
\end{cases}
\labl{Homotopies}
}
the only groups that have both these homotopy groups non-trivial
are $SO(n)$, $n=3,4,5$. However, the generators of $SO(n)$
are anti-symmetric hence the symmetric trace over the generators
($D_{ijk}$) is zero. The second possiblity is a non-trivial $\gp_5$
homotopy group
\equ{
\gp_5(G) = 
\begin{cases}
\Intr & SU(n \geq 3), \\
\Intr_2 & USp(2n).
\end{cases}
}
We see that the parity anomaly counter term may be gauge non-invariant
for all $USp(2n)$ groups and for $SU(n)$ qroups with $n \geq 3$.

Next, we consider the case that $G$ consists of several connected
factors. So let $G = G_1 \times G_4$ where the group $G_1$ is a
connected 
group and $G_4$ may contain various disconnected factors. By taking
different connected factors for $G_1$ and repeating the arguments
which are presented below, all possibilities can be found in this way. 
The field strength of the $G_1$
factor is denoted by $F_1$ and the field strength for the $G_4$ is
denoted by $F_4$. The relevant characteristic class $\gO_6$ then reads
(see \eqref{charCl})  
\equ{
\gO_6 = \Bigl( \frac i{2\gp} \Bigr)^3 \left[
\frac 16 \tr F_1^3 + 
\frac 12 \tr F_1 \tr F_4^2 + 
\frac 12 \tr F_1^2 \tr F_4 + 
\frac 16 \tr F_4^3 
\right].
}
We can disregard the first and last terms in this expression: the
first term we have already treated, while for the last term we simply 
repeat the analysis we are describing with $G = G_4$. From the
remaining two 6-forms we obtain four different 5-forms, but since they
are integrated over a manifold without boundary, we only need those
two of them ($\tr A_1 \tr F_4^2$ and $\tr F_1^2 \tr A_4$) which can be
readily integrated over $S^1\times S^4$. These terms can give
non-vanishing contributions to the gauge transformation 
of the parity counter term  
\equ{
 \gG_{PAC}({}^g A) \supset  
- \gp i\   \frac i{2\gp}\! \int_{S^1} \tr (g_1\inv \d g_1) \
 \frac {\,-1~\,}{8\gp^2}\!\! \int_{S^4}\tr F_4^2 
- \gp i\
\frac {\,-1~\,}{8\gp^2}\!\! \int_{S^4} \tr F_1^2 \ 
 \frac i{2\gp}\! \int_{S^1} \tr (g_4\inv \d g_4).
\labl{GaugeViolation}
}
We only have to focus on the first term, since the second term will be
taken into account when we take $G_1$ to be one of the other factors in $G$. 
Thus, for gauge violation of the parity counter term, we need that
both integrals of the first term are non-zero. 
The second integral gives the instanton number, which is 
classified by $\gp_3(G_4)$ which is $\Intr$ for any simple connected
Lie group. The first integral of the first term is the winding number 
of the gauge transformation that maps $S^1$ in the group $G_1$, 
classified by the fundamental group $\gp_1(G_1)$ given in
\eqref{Homotopies}. But as the generators of the $SO(n \geq 3)$ groups
are, traceless because of their anti-symmetry, only the
$U(1)$ can give rise to non-vanishing contributions. 

Since, we need gauge invariance for any background gauge field configuration, 
the instanton number is arbitrary. Gauge invariance of the parity 
counter term is also maintained if it changes by an integer multiple
of $2 \gp i$ as the effective action remains invariant.

Notice that if gravitational interactions are also taken into account, 
also the gravitational instanton can appear in \eqref{GaugeViolation}
so that also the gauge groups $G = U(1)$  can produce
a non-gauge invariant parity counter term. For this it is sufficient that 
\equ{
\int_{S^1} \frac i{2\gp} \tr (g_1\inv \d g_1) =  1\ \text{mod}\ 2,
}
for some large gauge transformations.

\end{document}